\pdfoutput=1
\documentclass{JINST}

\title{\bf Design and performance of a CMOS study sensor for 
a binary readout electromagnetic calorimeter}
\author{
J.\ A.\ Ballin$^a$,
R.\ Coath$^b$,
J.\ P.\ Crooks$^b$,
P.\ D.\ Dauncey$^a$\thanks{Corresponding author},
A.-M.\ Magnan$^a$,
Y.\ Mikami$^c$\thanks{Current address: Institut Pluridisciplinaire Hubert CURIEN, 23 rue du loess - BP28, 67037 Strasbourg CEDEX 2, France.},
O.\ D.\ Miller$^c$,
M.\ Noy$^a$\thanks{Current address: CERN, CH-1211 Gen\`eve 23, Switzerland.},
V.\ Rajovic$^c$\thanks{Current address: Faculty of Electrical Engineering, University of Belgrade, Bulevar kralja Aleksandra 73, 11120 Belgrade, Serbia.},
M.\ Stanitzki$^b$,
K.\ D.\ Stefanov$^b$\thanks{Current address: Sentec Ltd, Cambridge, UK.},
R.\ Turchetta$^b$,
M.\ Tyndel$^b$,
E.\ G.\ Villani$^b$,
N.\ K.\ Watson$^c$,
J.\ A.\ Wilson$^c$,
Z.\ Zhang$^b$\\
\llap{$^a$}Department of Physics,
Blackett Laboratory,
Imperial College London,
London, UK.\\
\llap{$^b$}STFC,
Rutherford Appleton Laboratory,
Chilton, Didcot, UK.\\
\llap{$^c$}School of Physics and Astronomy,
University of Birmingham,
Birmingham, UK.\\
\\
E-mail: \email{P.Dauncey@imperial.ac.uk}
}

\abstract{We present a study of a CMOS test sensor 
which has been designed, fabricated and characterised to
investigate the parameters required for a binary readout
electromagnetic calorimeter.
The sensors were fabricated with several enhancements
in addition to standard CMOS processing.
Detailed simulations and experimental results of the performance 
of the sensor are presented. The sensor and pixels are shown to 
behave in accordance with expectations and the processing enhancements
are found to be essential to achieve the performance required.
}


\begin{document}

\section{Introduction}
\label{sec:introduction}

In this paper, we present results from a test sensor
which was designed to investigate a novel approach
to sampling electromagnetic calorimeters (ECAL).
In this approach, the
readout of the sensitive layers of the ECAL are binary, 
meaning that each of the pixels which cover the
sensitive layers
return a yes/no result with no further amplitude 
information. With a fine enough pixel granularity, specifically
less than 100\,$\mu$m pitch such that
the probability of more than one charged particle hitting
a pixel is kept small, then the number of pixels reporting
a hit in a layer should be effectively a measure of the 
number of charged particles crossing that layer. In contrast,
an analogue readout ECAL measures the energy deposited by
the charged particles rather than counting them directly.
Hence, compared to the binary device,
the resolution of the analogue device will be degraded
by the fluctuations in the amount of energy deposited by
each charged particle. Therefore, a binary ECAL should
have an intrinsically better resolution for electromagnetic
showers.

Although the potential applications for a binary ECAL are
wide, the sensor discussed in this paper was designed to study
a specific application, namely the International Linear
Collider (ILC)~\cite{ref:introduction:ilc}. 
The work presented in this paper
has been done within the context of the
CALICE Collaboration~\cite{ref:introduction:calice},
which is studying calorimetry for future linear colliders.
For a recent status report on the collaboration activities,
see~\cite{ref:introduction:prc}.
One particular crucial aspect of ILC calorimetry is its
ability to support particle flow techniques~\cite{ref:introduction:pfa},
so as to obtain excellent hadronic jet energy resolution. These techniques
depend on spatial pattern recognition within the calorimeters.
A binary ECAL would have far superior spatial granularity (by
about two orders of magnitude) compared to an analogue ECAL
and so should also give better particle flow performance.

The basic motivation for a binary readout ECAL is
presented in~\cite{ref:introduction:decal} and a preliminary
comparison of an ECAL with a fixed sampling geometry but
with binary and analogue readout has been presented
recently~\cite{ref:introduction:ichep}. As pointed out in
the former, a critical limitation for any simulation
comparison of binary and analogue ECALs is that the
density of charged particles in the core of a high-energy
electromagnetic shower has not been measured at the
relevant granularities, namely less than 100\,$\mu$m.
Differences in the core density between data and simulation 
could have a significant impact on the achievable resolution of
the binary ECAL. Therefore, the sensor described in this
paper was developed for two purposes. Firstly, it allows a study
of the feasibility of implementing a MAPS sensor for a binary 
ECAL. Secondly, once a functional sensor has been developed, 
it can then be used to 
experimentally determine the electromagnetic shower core density at the
relevant granularity. This will
allow simulation to give an accurate and robust prediction of the 
performance of a full-scale binary ECAL.

This paper covers the first of these issues, namely the performance
of a MAPS sensor designed to study an architecture appropriate for a
binary ECAL.
In the rest of this paper,
the design and fabrication of the test sensor are 
presented in section~\ref{sec:design}.
Section~\ref{sec:simulation} describes the simulations
used for modelling the sensor response.
Sections~\ref{sec:testp} and \ref{sec:pixel} presents
results of individual pixel measurements, from test pixels
with analogue readout and bulk pixels with digital readout,
respectively.
Finally, section~\ref{sec:sensor} describes
results of sensor-wide measurements.

\section{Design and fabrication}
\label{sec:design}

In this paper, we present results from a test sensor
which was designed to investigate a binary ECAL
and to demonstrate the feasibility of this approach.
Because a complete ECAL
for this application would require a total
number of pixels of the order of $10^{12}$, this development has
been named the Tera-Pixel Active Calorimeter (TPAC) sensor.
This paper reports on results from the first two versions of a
test sensor, TPAC~1.0 and 1.1, which were based on a 0.18\,$\mu$m CMOS process.

The use of CMOS structures for light-sensing applications was originally 
proposed in the 1970s and CMOS imaging sensors were implemented in the
1990s~\cite{ref:design:history}.
Numerous recent developments to the 
pixel design and processing, such as the pinned photodiode, 
have achieved high-grade performance in areas such as 
low noise and leakage current. Such performance 
can be applied beyond the commercial imaging domain, and CMOS sensors have 
been demonstrated in many alternative applications including 
particle physics~\cite{ref:design:rt3}.

A key advantage of CMOS imaging is the use of standard CMOS fabrication processes, in which a diverse family of NMOS and PMOS transistors, resistors, capacitors and diodes can be manufactured on the same silicon substrate as the sensitive pixel; hence these sensors are often referred to as Monolithic Active Pixel Sensors (MAPS).  In typical imaging applications this allows 
the integration of the readout electronics and control logic at the edge of the array of imaging pixels, offering a very compact one-chip imaging solution.  
Although the requirements for particle physics applications of CMOS sensors share 
many aspects with those for commercial imaging, such as low noise performance, 
the pixel pitch needed is usually larger. For the binary readout ECAL,
pixel sizes of order
100\,$\mu$m are needed~\cite{ref:introduction:decal}. 
This large pixel area provides an opportunity to exploit the full range of 
components 
available in the CMOS process, with the possibility to add significant
amounts of circuitry inside these larger pixels, rather than around the
periphery of the sensor. While these in-pixel components would reduce
the light reaching the sensor for imaging applications, they have no 
effect on the passage of charged particles.

\subsection{Process technology}
\begin{figure}[ht!]
\begin{center}
\includegraphics[width=16cm]{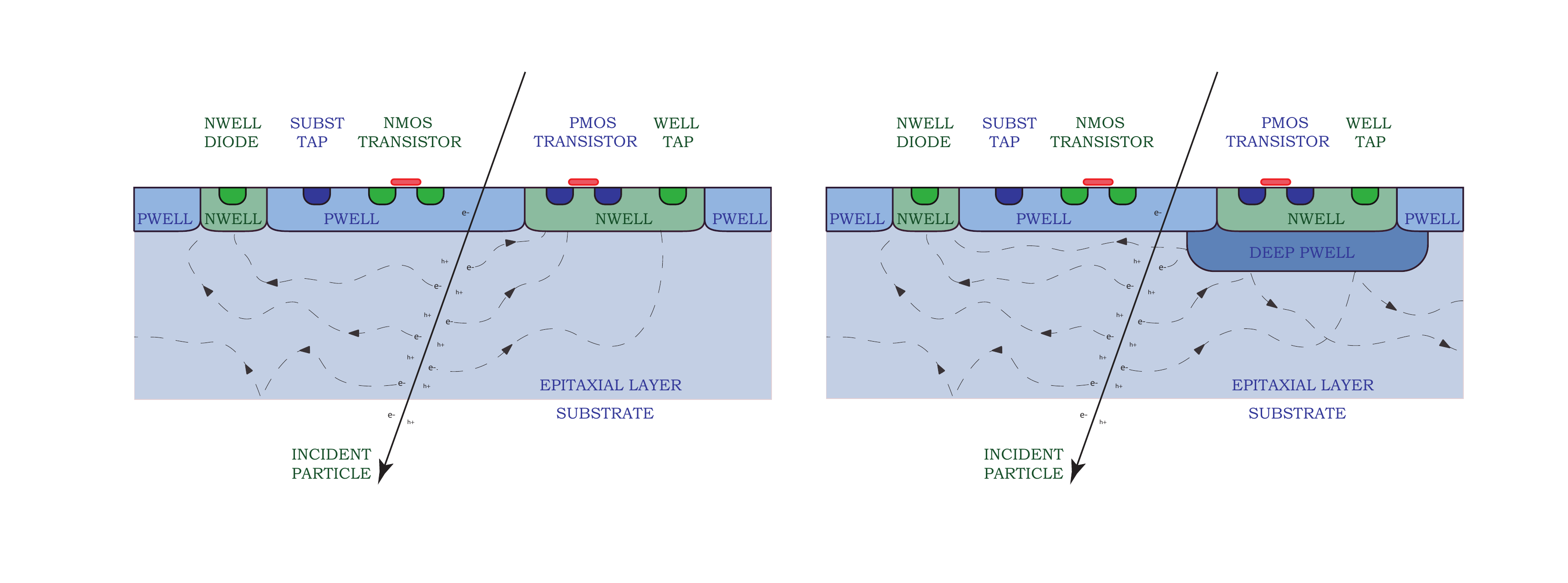}
\caption{\sl 
Conceptual illustration of a
charged particle crossing a CMOS sensor epitaxial layer.
The N-well to P-epitaxial diode at the upper left is the
signal collecting diode.
The one at the upper right encloses a PMOS transistor which is
part of the in-pixel circuitry. Left: the latter is exposed and
so can absorb signal charge, reducing efficiency. Right: the N-well
diode is shielded from
absorbing signal charge by the deep P-well implant between it
and the epitaxial layer, reducing signal loss.
}
\label{fig:design:cmos}
\end{center}
\end{figure}

The CMOS silicon substrate comprises a very low resistivity base material, 
over which a P-doped epitaxial layer is grown, generally up to 20\,$\mu$m 
in thickness.  
For the application described here, epitaxial layer thicknesses
of 5 and $12\,\mu$m were used.
The silicon substrate and epitaxial layer are predominantly free of 
electric fields, so any charge that has been deposited in the silicon 
will move randomly by diffusion, with typical carrier lifetimes of 
milliseconds, and will be collected in times of the order of 100\,ns
(see section~\ref{sec:testp}).
A potential barrier exists due to the change in substrate doping 
between the silicon substrate and the epitaxial layer as well
as between the heavily doped P-wells and the epitaxial layer, 
which is sufficient 
to keep the majority of carriers within the epitaxial layer.
Also an electric field forms around the positively biased N-well to
P-epitaxial diode, 
in which a diffusing electron will be swept towards the N-well and 
collected as signal.
Figure~\ref{fig:design:cmos} illustrates the basic principles.

As stated 
above, the epitaxial layer is the detecting volume where charge
is generated.
A good 
approximation is to consider that the most probable 
total number of electron-hole 
pairs generated by a MIP in silicon is equal to 80 pairs/$\mu$m.
For the 12\,$\mu$m thick 
epitaxial layer used in TPAC~1.0, this corresponds to only about 1000 
electron-hole pairs. 
There will also be some contribution 
to the charge generation from the upper few microns of the substrate
which increases the MIP signal charge to around 1200\,$e^-$. 
However, given that the charge can freely diffuse between pixels, 
the number of charge carriers collected by any single pixel is
smaller. Clearly, 
any further loss, specifically due to charge collection by 
unrelated N-wells, would make the efficient
detection of MIPs in CMOS sensors
very difficult, if not impossible.

Due to diffusion, the signal charge will not necessarily be collected by the 
nearest diode, in which case it continues to diffuse and may be collected 
by a diode in a neighbouring pixel.
In particle physics tracking applications or in imaging sensors
this effect is called ``crosstalk'' and represents an undesirable loss of 
image quality for the latter.
Although this is less of an issue for an ECAL sensor, it will
still degrade the performance in this application and so
the pixel design presented in this paper implements four diodes, 
placed toward the corners of the pixel for optimum crosstalk reduction.
The advantage of this approach is also a reduction in charge collection 
time, as the mean distance to any diode is shorter than the single-diode 
solution at the same pixel pitch.

The structure of a standard PMOS transistor presents difficulties for
use in a pixel design, due to the N-well in which the PMOS device sits.
Such N-wells, tied to a positive potential, are as likely to collect 
diffusing charge as the collecting diode. However, any charge which reaches 
these N-wells will not be collected as signal on the diode, and so this 
behaviour represents an inefficiency in charge collection.
For this reason, commercial CMOS sensors normally only use NMOS transistors 
within the pixel, although this significantly limits the circuit functionality 
that can be implemented.
For a complex pixel design with many PMOS transistors and small collecting 
diodes, this inefficiency could dominate the charge collection and the 
resulting signal size would be too small to resolve over the electronic noise.

To address the problem of inefficiency caused by PMOS transistors in the pixel,
we proposed adding a deep P-well implant to the standard CMOS process.
This high-energy implant creates a heavily P-doped 
beneath the N-well of a PMOS transistor, as shown in 
figure~\ref{fig:design:cmos}.
The potential barrier that forms, much like the boundary between the 
epitaxial layer and the substrate, is again sufficient to keep the majority 
of carriers within the epitaxial layer, and more importantly, from being 
collected by the N-well.
This technique restores the charge collection efficiency of the sensor
to be
close to 100\%, although some minor reduction in the initial signal charge 
will occur, as charge deposited in the small proportion of the epitaxial layer 
that is now occupied by the deep P-well implant will be lost.

For the successful implementation of this project, the deep P-well module was 
implemented in a leading 0.18\,$\mu$m commercial process.  
This fabrication process is called ``Isolated N-well MAPS'' (INMAPS)
and is described
in detail elsewhere~\cite{ref:design:inmaps}.  
The INMAPS process features up to 
six metal layers, precision passive components 
for analogue circuit design, and may be stitched to manufacture sensors up 
to wafer scale.  Whilst for this project a particular commercial partner was 
selected, the technology could be implemented in many modern CMOS processes.
This is an important feature if large-scale production were ever required
for a full-size calorimeter.

The TPAC sensors were manufactured using the INMAPS process, and implemented
deep P-well implants
in the pixels to achieve a high charge collection efficiency.
To further understand the device physics, the same design was manufactured 
with two thicknesses (5\,$\mu$m or 12\,$\mu$m) of the epitaxial layer and 
with or without the deep P-well implant.

\subsection{Overall architecture}
The TPAC sensors comprise a $168\times 168$ pixel array, totalling
28,224 pixels, together with row control logic, on-sensor
SRAM memory banks and I/O circuitry in a $9.7\times 10.5$\,mm$^2$ die.
The sensor has been developed with the ILC application in mind. It
is designed to be sensitive during a ``bunch train'', a short
period of the order of 1\,ms, 
which is subdivided into ``bunch crossings'', typically
with several thousand bunch crossings per train.
This structure is consistent with the proposed ILC operation.
The sensors collect the charge deposited by an incident particle in pixels 
arranged on a 50\,$\mu$m pitch.
This signal is compared with a global threshold and if above threshold,
the bunch crossing time-code and location of the 
pixel ``hit'' is recorded in on-sensor memory for readout 
following the bunch train.

Four different pixel designs have been implemented for evaluation, 
which fall into two distinct architectures, called pre-shape and pre-sample. 
These four designs are arranged in four quadrants of $84 \times 84$
pixels.
All pixels contain four small N-well diodes for charge collection
and these are identical for all four designs.
Pixels may be individually masked, allowing any permutation of single 
pixels to be operated and evaluated.

\begin{figure}[ht!]
\begin{center}
\includegraphics[height=5cm]{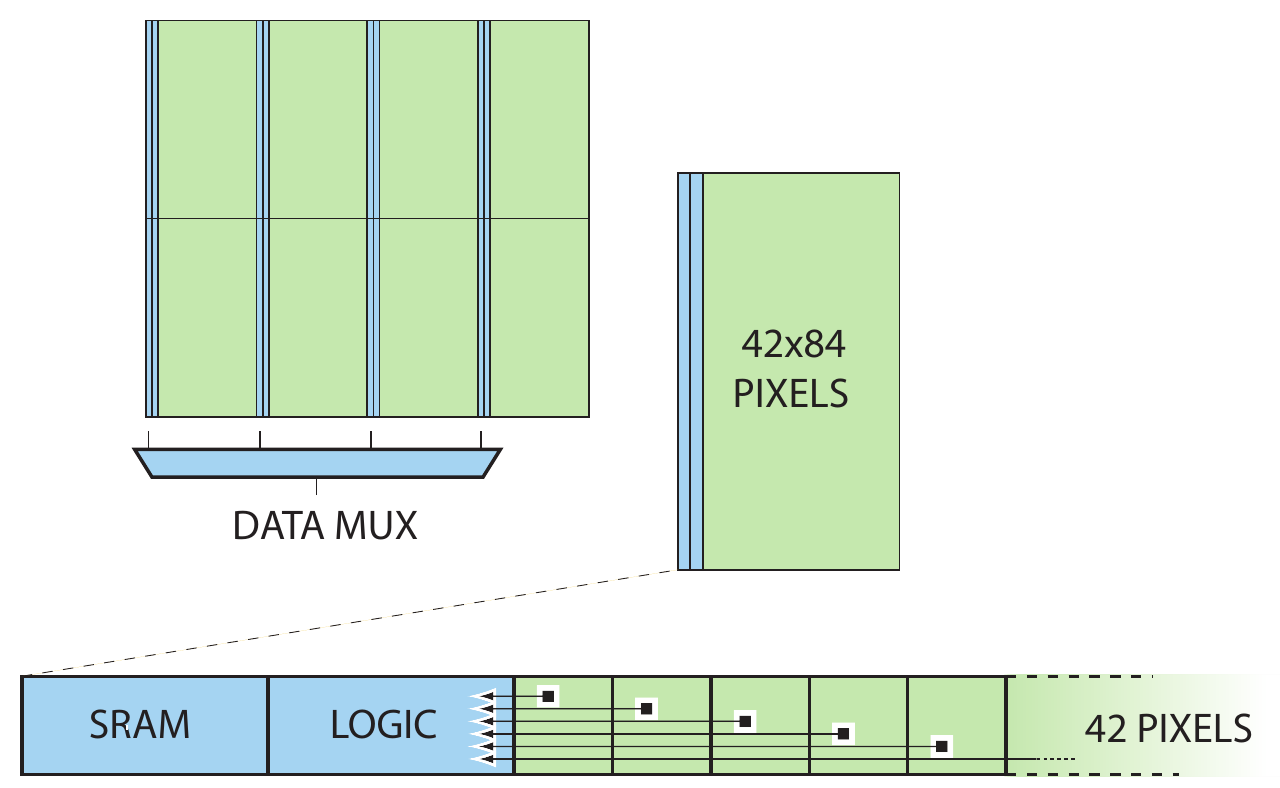}
\includegraphics[height=5cm]{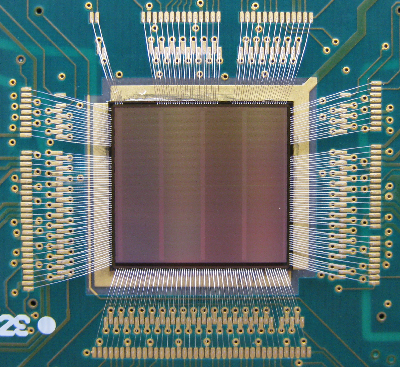}
\caption{\sl 
Left: overview of sensor architecture; see text for a
detailed explanation.
Right: photograph of the sensor mounted on its PCB.
The four pixel regions and memory areas are visible within
the central area of the sensor.
}
\label{fig:design:architecture}
\end{center}
\end{figure}

A common control and readout architecture serves all pixel varieties, 
allowing the sensor to be operated as a whole or as sub-regions.
Rows of 42 pixels share row control logic and SRAM memory, while
columns of 168 such rows form a region which shares data readout.
All four regions multiplex their data for external readout off-sensor.
An overview of the architecture is shown in 
figure~\ref{fig:design:architecture}, together with a photograph of
the TPAC~1.0 sensor.

\subsection{Pre-shape pixel}
\label{sec:design:preshape}

\begin{figure}[ht!]
\begin{center}
\includegraphics[width=14cm]{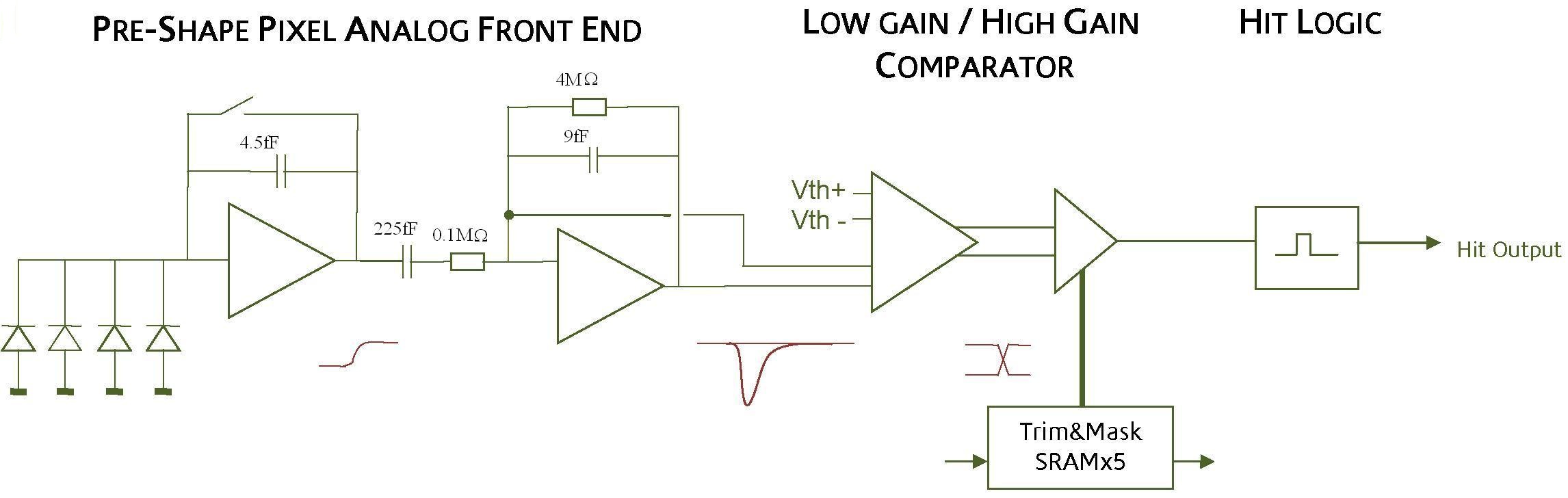}
\caption{\sl 
Pre-shape pixel circuit; see text for a detailed explanation.
}
\label{fig:design:preshape}
\end{center}
\end{figure}

The pre-shape pixel is based on a conventional analogue front end for a 
charge-collecting detector and is shown in figure~\ref{fig:design:preshape}.

\begin{figure}[ht!]
\begin{center}
\includegraphics[width=7cm]{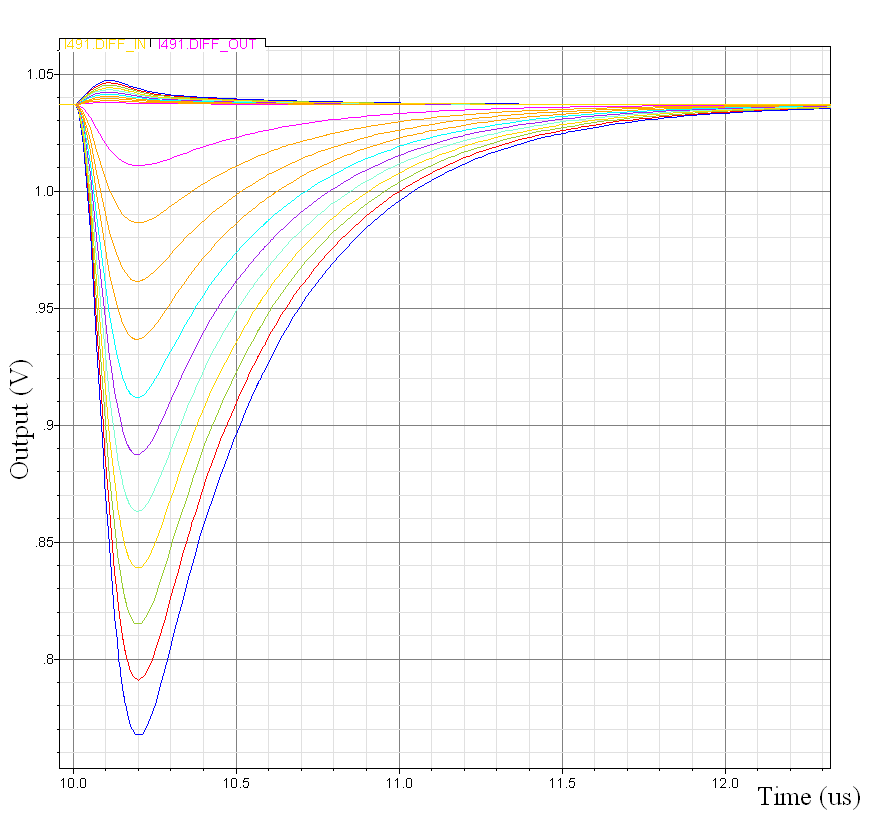}
\caption{\sl 
Simulated pre-shape pixel circuit response as a function of time, for
signals injected at the preamplifier input at
a time corresponding to
10\,$\mu$s on the time axis. 
The two pseudo-differential comparator inputs are shown. The small
positive-going traces are from the preamplifier output, which provides an
effective common mode DC reference that
tracks process variations. The larger
negative-going traces
are from the shaper output, and these give the signal.
The input signals shown 
have various magnitudes up to a maximum of
2500\,$e^-$, in steps of 250\,$e^-$. The input signal shape is
modelled as an exponential rise with a 50\,ns time constant.
Together with the finite circuit response speed, this
means the observed signals peak around 200\,ns after the signal
injection time.
}
\label{fig:design:preshapemips}
\end{center}
\end{figure}

Four diodes are connected to a charge preamplifier, which generates a voltage 
step output in proportion to the collected charge.
A CR-RC shaper circuit with time constants around 30\,ns
then generates a pulse with 
output size in proportion to the input signal.
However, the shaper circuit response is constrained by the slower preamplifier 
output rise time, so
the overall peaking time for a impulse input is around 150\,ns.
Including the further circuit gain, full circuit simulation predicts a gain
of 164\,$\mu$V/$e^-$
with respect to total input charge.
The simulated response of the charge preamplifier and shaper,
as well as a local common-mode reference,
to signals of varying
magnitude is shown in figure~\ref{fig:design:preshapemips}.
These signals form the
pseudo-differential inputs to the two-stage comparator.  

The preamplifier reset is used to initialise the pixel before the start of a 
bunch train but is then held open during data-taking. During the train,
following a charge deposit,
the shaper circuit returns to a stable state after a time
depending on the signal size,
and is then able to respond to another input deposit.
Saturation in the shaper circuit occurs for individual
signal charge deposits greater 
than 2500\,$e^-$, beyond which the shaper output becomes non-linear and takes 
a longer time to return to the steady state.
In addition, because the preamplifier is not
reset during the bunch train,
saturation of the pixel will also occur if the preamplifier 
stage integrates 10\,k$e^-$ in total on the diode node during a bunch train. 
Beyond this, the gain 
of the pixel deteriorates, reaching 50\% after 22\,k$e^-$,
until it will no longer respond to an input signal.
The expected incident signal charge for a single pixel during a bunch
train for the ILC ECAL application is small compared to these saturation 
limits.

To achieve high circuit gain in the preamplifier, a small feedback 
capacitance of 4.5\,fF was required (see figure~\ref{fig:design:preshape}), 
which was made using two larger 
capacitors in series 
to comply with manufacturing design rules.
Two different simulation tools, Eldo~\cite{ref:design:eldo} and 
Spectre~\cite{ref:design:spectre},
were used to evaluate the optimum orientation 
of the series feedback capacitors, but the two tools selected different 
topologies for highest gain.
Two capacitor orientations were therefore implemented on the TPAC~1.0 
sensor as 
subtle variants of the pre-shape pixel and these occupy quadrants~0 and 1.
Results from and comparisons of the performance of both quadrants~0 and 1 are
shown in the following sections.

The in-pixel comparator has two stages. The first stage takes two differential 
signals, and produces a real-time differential result, 
with some small analogue signal gain.
The second, high-gain comparator generates the full-swing 
discriminator output, 
and applies an offset trim adjustment with a four-bit resolution.  
The output of the second comparator is enabled with a one-bit mask
which can 
be used to prevent the pixel from generating hit outputs.

Pixel signals which cross through the threshold from below
generate a fixed length pulse using a monostable circuit, the output of
which is connected to row control logic outside the pixel.
The length of the resulting hit output pulse is thus
independent of the signal size.

Electronic circuit noise is estimated at the input to the differential 
comparator and referred back to the diode node using the charge gain.
The dominant noise source is the input transistor of the preamplifier 
circuit; the predicted equivalent noise charge (ENC) for this pixel is
23\,$e^-$.

The nominal power consumption for the pre-shape pixel is 8.9\,$\mu$W during 
operation, although it is assumed that in any realistic sensor, the circuit 
would be powered off between bunch trains at the ILC. This mode of operation
was not supported by the TPAC system described here so it has not been tested.
However, other studies within CALICE have demonstrated that this is a
feasible assumption~\cite{ref:introduction:prc}.
Assuming a factor of 100 reduction in power by this method, this is
equivalent to an average power consumption of 36\,$\mu$W/mm$^2$
and hence is is significantly larger than the target power of the analogue 
ECAL designs for the ILC detectors~\cite{ref:design:sid, ref:design:ild}.
However, the main aim of the TPAC~1.0 sensor is to 
investigate the concept of a binary ECAL and demonstrate the feasibility
of the design with a new technology.
Power was regarded as
a secondary issue, at least for this first demonstrator.
Lower power designs will be considered in future versions of the sensor.

\subsection{Pre-sample pixel}
\label{sec:design:presample}

\begin{figure}[ht!]
\begin{center}
\includegraphics[width=14cm]{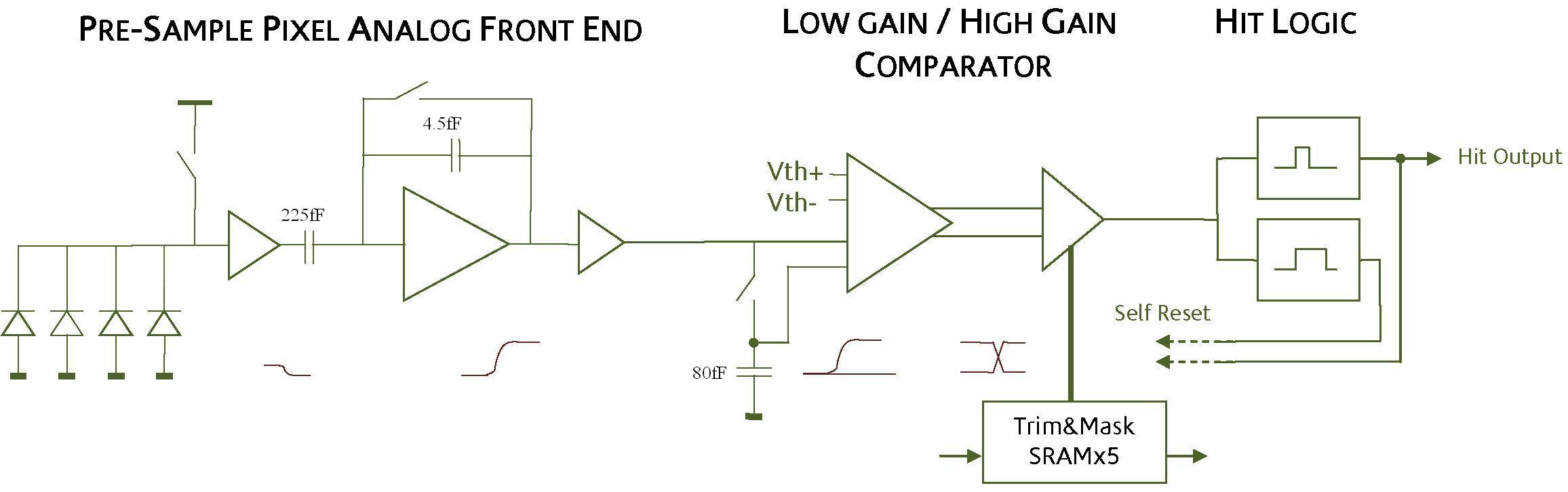}
\caption{\sl 
Pre-sample pixel circuit; see text for a detailed explanation.
}
\label{fig:design:presample}
\end{center}
\end{figure}

The pre-sample pixel architecture is closer to that of
a conventional MAPS optical sensor. It has
in-pixel analogue storage of a reference level, and is 
shown in figure~\ref{fig:design:presample}. This design was found
to perform worse than the pre-shape circuit in initial tests
and testing was stopped
early on to concentrate on the latter. 
Therefore, no detailed study of its performance was done,
and hence no pre-sample pixel
results are shown in the rest of this paper.
Only pre-shape pixels were
implemented in the next version of the sensor, but
the pre-sample design is described here for completeness.

The pixel diodes are reset and the
reset level is sampled and stored in-pixel
prior to the start of each bunch train.
Charge integrates on the four collecting diodes, causing a small voltage 
step proportional to the collected charge and the node capacitance.
A charge preamplifier provides a
voltage step which, along with the local sample of the reset level, 
forms a pseudo-differential input to the same two-stage comparator
as for the pre-shape design.
The predicted gain of the circuit is 440\,$\mu$V/$e^-$.
The charge amplifier and reference sample must be reset after a hit 
before the pixel can detect another hit;
this is undertaken by the in-pixel logic.    

\begin{figure}[ht!]
\begin{center}
\includegraphics[width=7cm]{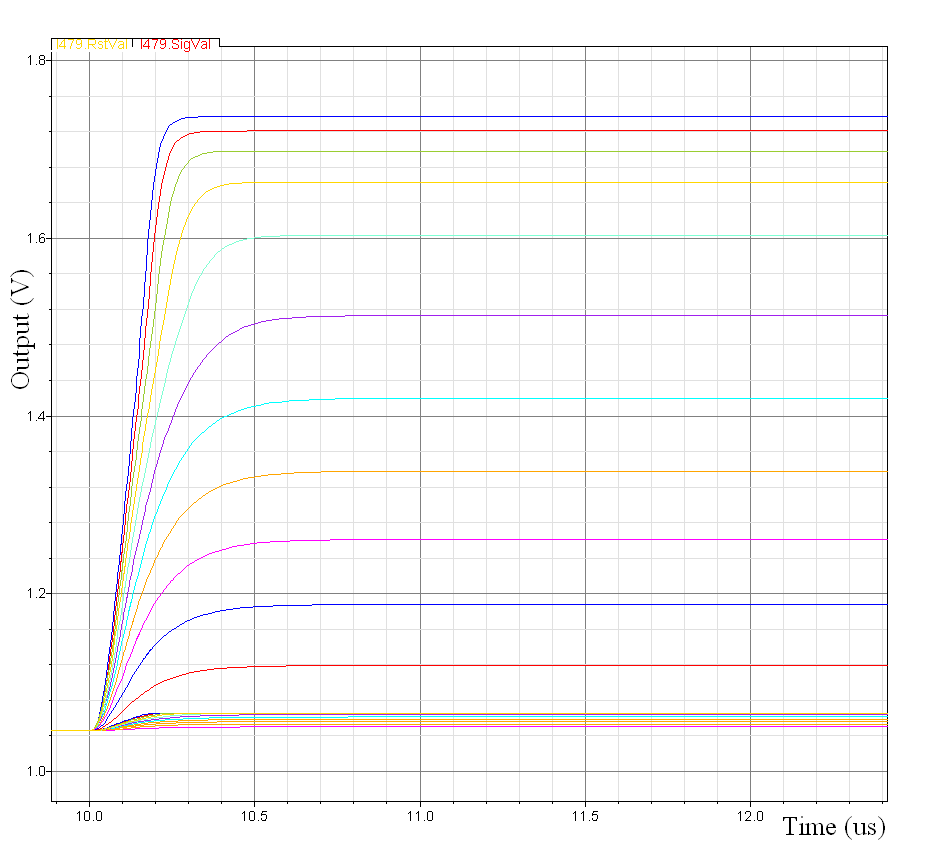}
\caption{\sl 
Simulated pre-sample pixel circuit response as a function of time to input
signals injected at 
a time corresponding to
10\,$\mu$s on the time axis. 
The input signals shown 
have various magnitudes up to a maximum of
2500\,$e^-$, in steps of 250\,$e^-$. The input signal shape is
modelled as an exponential rise with a 50\,ns time constant.
}
\label{fig:design:presamplemips}
\end{center}
\end{figure}

The simulated response of the circuit to signals of varying
magnitude is shown in figure~\ref{fig:design:presamplemips}.
Saturation in the pre-sample pixel occurs when the diode node has integrated 
64\,k$e^-$, beyond which non-linear operation is expected.

Similarly to the pre-shape pixel, a small capacitance in the preamplifier 
feedback is made with two capacitors in series.
This gives rise to two subtle variants of the pre-sample pixel, 
again based on results from different simulator tools, and these
occupy quadrants 2 and 3.

The in-pixel comparator stage is common to all pixel architectures, 
but the pre-sample pixel includes an additional monostable circuit 
to generate the self-reset signals that are necessary to reset the 
amplifier and the reference sample in preparation for another input
signal.

Electronic circuit noise is estimated at the input to the differential 
comparator and referred back to the diode node using the charge gain.
The circuit simulation predicts an ENC at this input of 22\,$e^-$. 
The dominant noise sources are the input transistor of the first source 
follower buffer and the input device in the charge amplifier circuit.
However, the overall observed noise when operating the pixel will be
larger by an additional factor of $\sqrt{2}$ 
to account for the uncorrelated sampling nature of this pixel.
This correction needs to be applied to the real-time simulation 
noise level and so predicts an overall noise of 37\,$e^-$.

The nominal power consumption for the pre-sample pixel is 9.3\,$\mu$W 
during operation, although again the circuit would be powered off between 
bunch trains at the ILC.



\subsection{Mask and trim configuration}
\begin{figure}[ht!]
\begin{center}
\includegraphics[width=6cm]{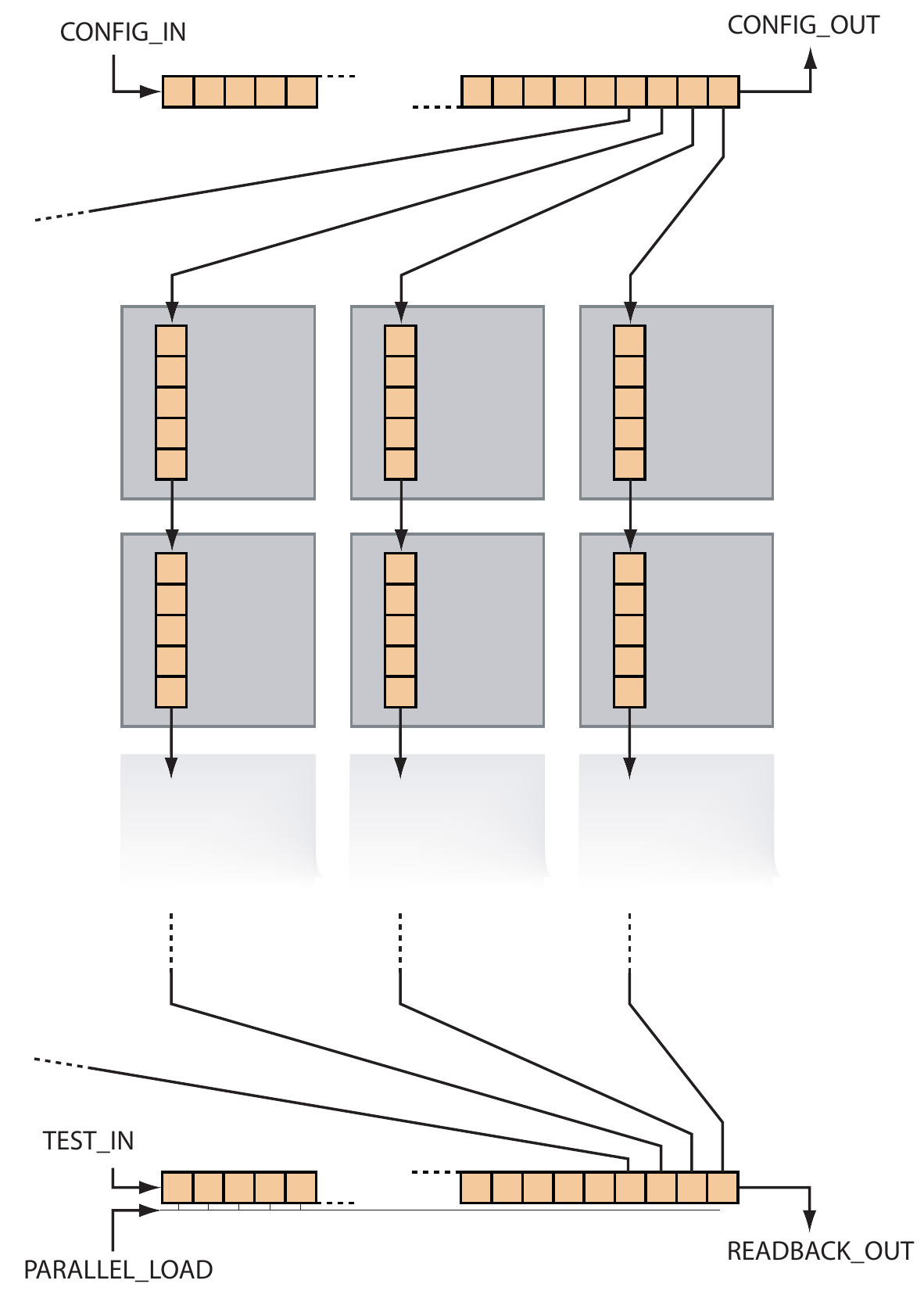}
\caption{\sl 
Overview of logic for configuration data write and readback.
To load, one bit of the configuration data for each column are shifted 
into the register at the top of the diagram. When completed, these
bits are parallel-loaded into the pixels below. This is continued
until all pixel bits are loaded. The previously stored configuration
data are captured in the register at the bottom of the diagram and
read out, allowing a consistency check.
}
\label{fig:design:config}
\end{center}
\end{figure}

Each pixel contains a five-bit SRAM shift register, 
which is used to store a per-pixel trim (four bits) and a mask flag (one bit).
The configuration shift registers are not used while the sensor 
is in active operation,
but these configuration data are programmed during sensor initialisation,
and are held indefinitely in each pixel until the sensor is powered down 
or the data are rewritten.
The configuration data are loaded through a serial interface, 
which shifts single-bit data into each of 168 columns simultaneously,
as illustrated in figure~\ref{fig:design:config}.
Serial data outputs are available at two points to enable data read-back 
for error-rate monitoring.
The data read is destructive, so in normal operation the read-back occurs 
after hits have been collected.
The total configuration memory space on the sensor is 138\,kbits.

\subsection{Pixel layout}
\begin{figure}[ht!]
\begin{center}
\includegraphics[width=6cm]{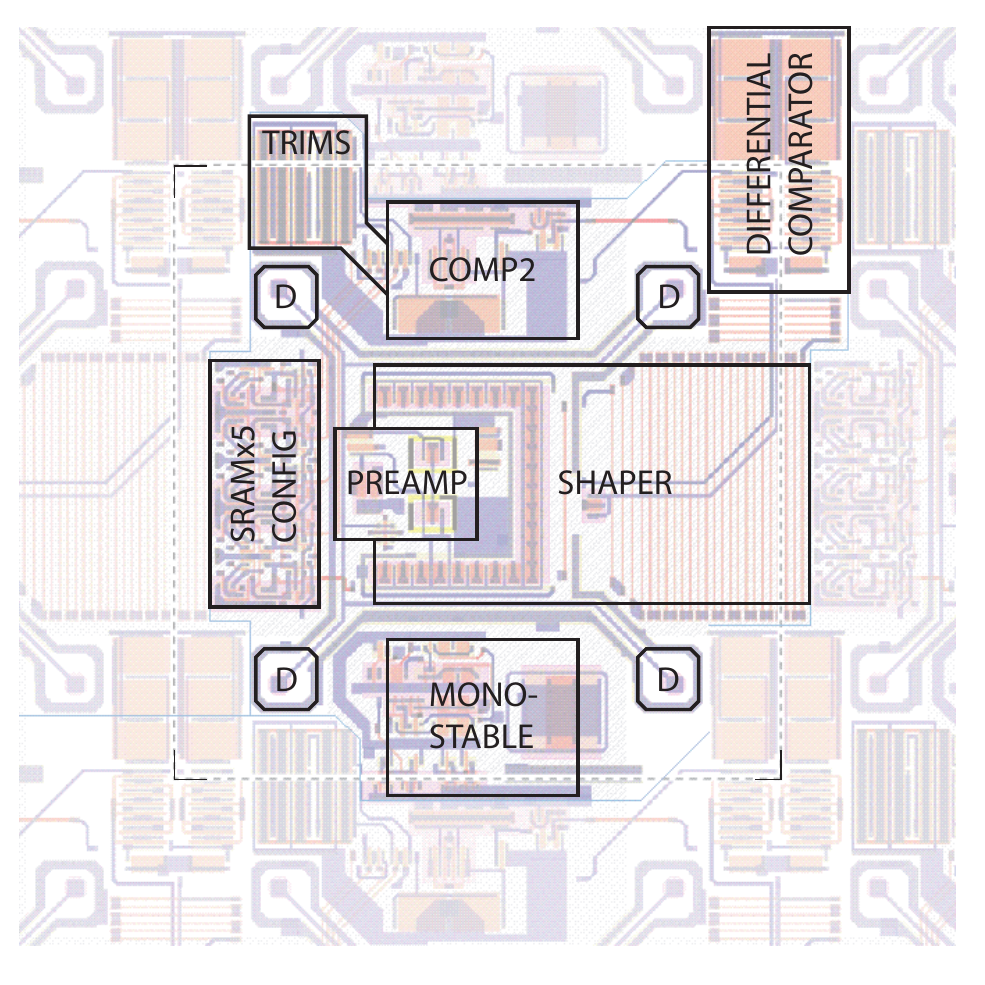}
\phantom{XXX}
\includegraphics[width=6cm]{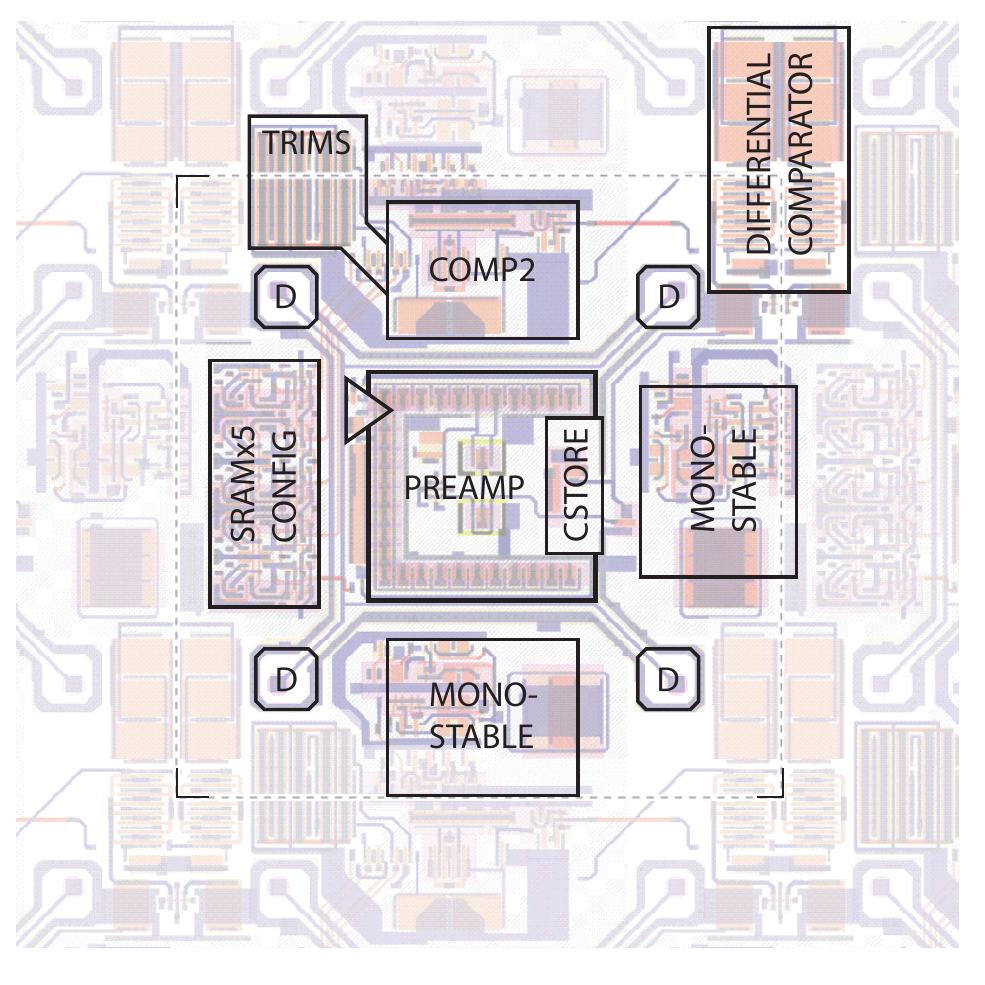}
\caption{\sl 
Pixel layouts; pre-shape (left) and pre-sample (right).
The boxes labelled ``D'' indicate the diode positions.
The nominal borders of the pixels are shown by the
right-angled corners and dotted lines.
}
\label{fig:design:layout}
\end{center}
\end{figure}

The pre-shape and pre-sample pixel layouts are illustrated in 
figure~\ref{fig:design:layout}.
The pre-shape pixel comprises 160 transistors, 27 capacitor cells and a 
large polysilicon resistor.
The pre-sample pixel comprises 189 transistors and 34 capacitor cells.
The two pixel architectures use the same diode positions.
The diodes are not located exactly on a 25\,$\mu$m pitch square grid but
were moved slightly towards
the corners of the 50\,$\mu$m pixel. These positions were 
optimised for charge collection using the simulation
described in section~\ref{sec:simulation:sentaurus}.

\begin{figure}[ht!]
\begin{center}
\includegraphics[width=6cm]{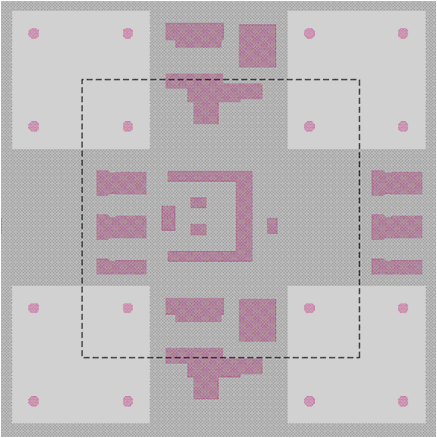}
\phantom{XXX}
\includegraphics[width=6cm]{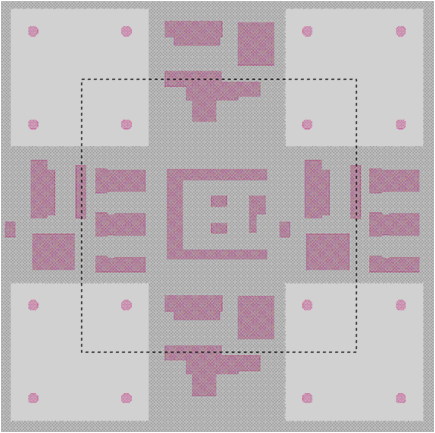}
\caption{\sl
Pixel deep P-well implant layouts; pre-shape (left) and pre-sample (right).
The orientation is the same as in the previous figure.
The nominal border of the pixel is shown by the dashed lines, so the areas
outside this are part of the eight neighbouring pixels.
The grey shows the deep P-well implant area while the purple shows
the N-well structures. All the N-well regions are protected by deep P-well
implant except for the four signal diodes towards the pixel corners.
}
\label{fig:design:dpwLayout}
\end{center}
\end{figure}

The sensitive analogue front-end circuits are located in the very centre 
of the pixel with extensive substrate-grounded guard rings for signal 
integrity.
Analogue signals are routed primarily on metal layer 1, 
with some plates of metal layer 2 used where necessary to shield the analogue 
signals from switching signals passing overhead.
The deep P-well layer is added as a symmetrical cross structure which leaves 
only the collecting N-well diodes exposed to the epitaxial layer,
as shown
in figure~\ref{fig:design:dpwLayout}.

Pixel power supplies are routed on the metal layers 3, 4 and 6 
in horizontal and vertical directions to distribute power in a mesh structure.
The various sub-circuits in each pixel design are mostly powered separately in 
this first sensor, so there are five independent power supplies routed to the 
pixels.
The hit output signals from pixels are routed horizontally along a row on 
metal layer 5, which is fully shielded from the sensitive analogue electronics 
below by the metal layers in between.

\subsection{Row control logic}
The row logic is responsible for monitoring the individual hit outputs from 
a row of 42 pixels and writing details of any hits to local memory.
An external clock defines the timing with which hit signals are sampled.
The hit signal from a pixel is asynchronous, but will have a fixed output pulse
width defined by the in-pixel monostable bias setting.
This pulse length would typically be set to be around 10\% greater than the 
hit sampling period,
which would be matched to the bunch crossing rate of the target 
application, typically 200-400\,ns.
This regime ensures that an asynchronous hit will always be sampled 
by the synchronous logic, with a small probability that it will be sampled 
twice.
This is an acceptable data overhead that allows for a reasonable spread 
in the length of the monostable pulses, with a minimal risk that an entire 
hit pulse occurs between sampling such that the hit is lost.
The sampling of hits uses an alternating (``ping-pong'') circuit 
architecture to ensure 
there is no dead time between samples.

\begin{figure}[ht!]
\begin{center}
\includegraphics[width=10cm]{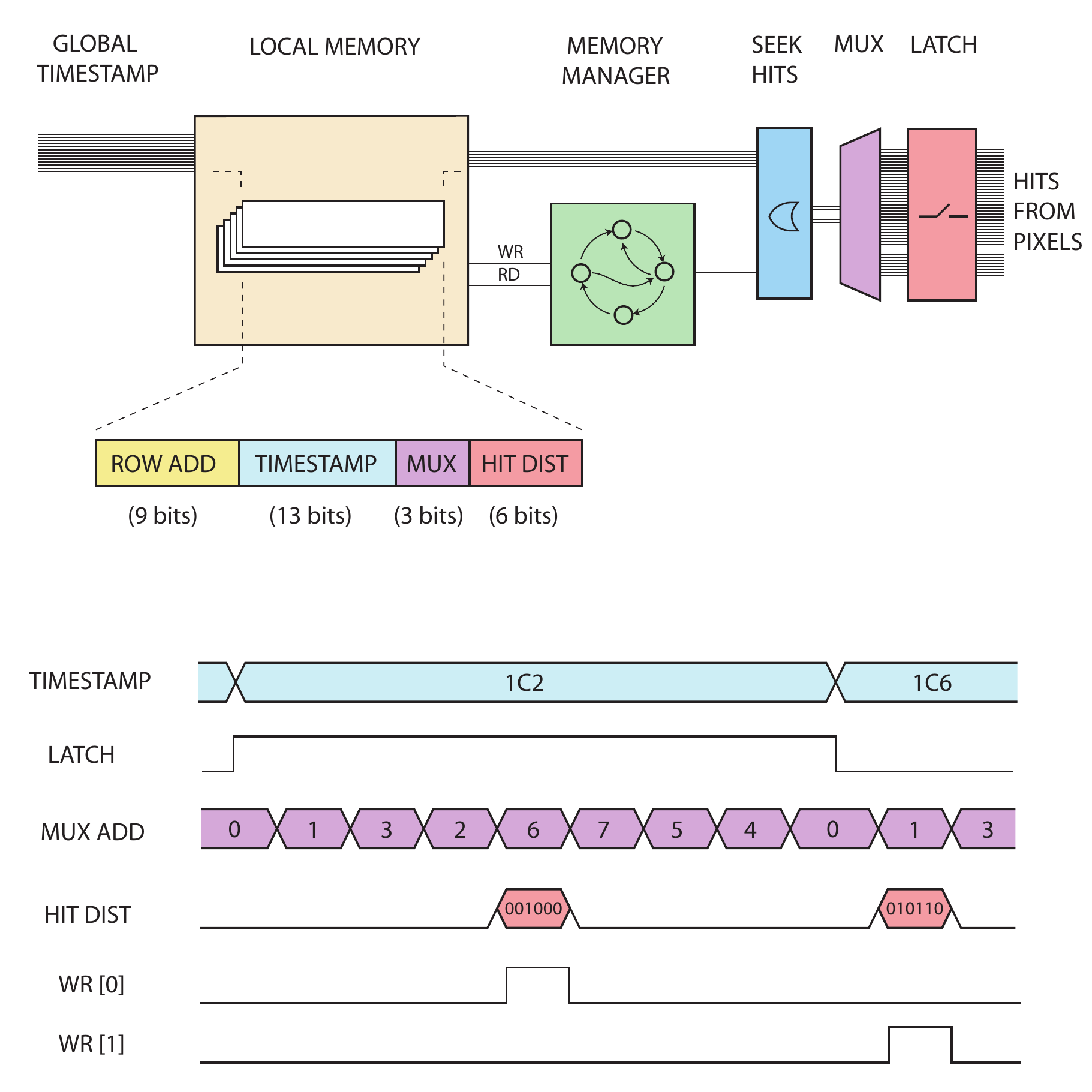}
\caption{\sl 
Overview of the row logic for hit storage; see text for details.
}
\label{fig:design:rowlogic}
\end{center}
\end{figure}

The 42 sampled hit signals are subdivided into seven ``banks'' for most 
efficient processing and storage.
Each bank is selected for inspection
in turn with a three-bit multiplex (MUX) address code.
An OR circuit then tests the six pixel outputs of the bank to see
whether the bank contains any hits.
If so, the six-bit pixel hit pattern and the three-bit MUX
address code are stored,
thus identifying a single location in the full row of 42 pixels.
This is illustrated in figure~\ref{fig:design:rowlogic}.
A feature of this approach is in the case of a dense particle shower, 
for which multiple nearby hits are stored in a single register, rather than 
multiple registers.
The hit-seeking circuits operate at eight times the bunch crossing rate, 
i.e. up to 50\,MHz.
Multiplex signals are gray-coded for reliable high speed operation, 
and the reserved address value 0 deselects all banks for additional testing 
provision. 

\subsection{Data storage}
\label{sec:design:storage}

The row control logic has 19 SRAM registers available for storage of hit data.
A memory controller is implemented to organise the use of these registers, 
such that registers are not overwritten once used, and only those with valid 
data participate in readout.  This memory controller is implemented as a 
bidirectional shift register, with 20 cells.  The memory control register 
initialises with a token in the first position, which enables the first SRAM 
register for write access.  During data write the shift register advances to 
the next position, filling the SRAM registers in order for the first 19 hits.
A further hit in the row then moves the token into a holding cell, 
which raises a global overflow flag indicating the memory full status.
This hit, and any further hits on that row, will be discarded.  

The row control logic may be operated in ``hit override'' mode, whereby the 
result of the OR circuit is ignored and the value of the hit pattern in each 
bank is always stored.  This operating mode fills the memories in less than
three complete cycles of the standard control sequence, and so is intended as 
a test feature.

The 19 SRAM registers occupy the full 50\,$\mu$m row pitch.  
The hit pattern and corresponding MUX address are stored in the first 
9 bits of a register, with a further 13 bits used to store the global 
timestamp code, which is incremented each time hit signals are sampled.
The cross-coupled inverter structure of a SRAM cell ensures the data will 
be held indefinitely provided the cell is powered, so there is no requirement 
to refresh the data and no maximum hold time after which data are corrupted.  

The full TPAC~1.0 sensor comprises four columns of row logic, 
each with 168 rows, hence there are 12,768 SRAM registers of 22 bits each 
in total.
The row control logic and the SRAM register bank occupy a 250\,$\mu$m wide 
region adjacent to the 42 pixels, equivalent to the area of five pixels.
There is no sense circuitry in this region
and, as no deep P-well is added 
here, charge arising from particles that pass through these regions 
will be collected by local N-wells associated with PMOS transistors
in the logic and SRAM.
This charge will not be collected as signal and
the logic and SRAM are therefore insensitive to incident particles.
In addition, an inactive 50\,$\mu$m wide strip
across the centre of the sensor 
corresponding to a single row of pixels,
is used to distribute bias and reference voltages, and to 
re-buffer control signals.
These inactive regions
result in the sensor having an inherent 11\% dead area.

\subsection{Readout}
During readout, the memory controller is switched into the reverse direction 
and clocked once to initialise the token to enable the most recently written 
register for readout.  
A combinational read-enable signal propagates to the first register that has
valid data for readout, and enables the connection to the parallel readout bus.
On each subsequent clock of the memory controller the next valid register is 
selected until no further registers remain, at which point a ``done'' output 
flag is asserted.
The off-sensor control software uses this flag to initiate readout of each 
logic column in turn.

In addition to the 22-bit SRAM registers, a 9-bit ROM cell is activated during 
readout of each row.  These extra cells encode a unique row address that forms 
part of the parallel data bus.

The SRAM and ROM readout is implemented with a current sense amplifier
which was optimised to operate over long distances with minimal read time.
An activated SRAM or ROM cell pulls current from the parallel data bus 
depending on its state, which is sensed by the circuit at the column base.
A total of 31 of these current-sense amplifiers operate in parallel 
to create a 31-bit digital output data word,
which is multiplexed and driven off-sensor with no 
serialization.
The maximum read time from the furthest cell is 150\,ns and
parallel data readout is typically operated at a 5\,MHz rate.
A full sensor readout, in which every register contains valid hit data (such 
as when operating in hit override mode) 
therefore takes approximately 2.6\,ms, and generates 50\,kBytes of data.

\subsection{Pixel test structures and additional test features}
\label{sec:design:test}

\begin{figure}[ht!]
\begin{center}
\includegraphics[width=14cm]{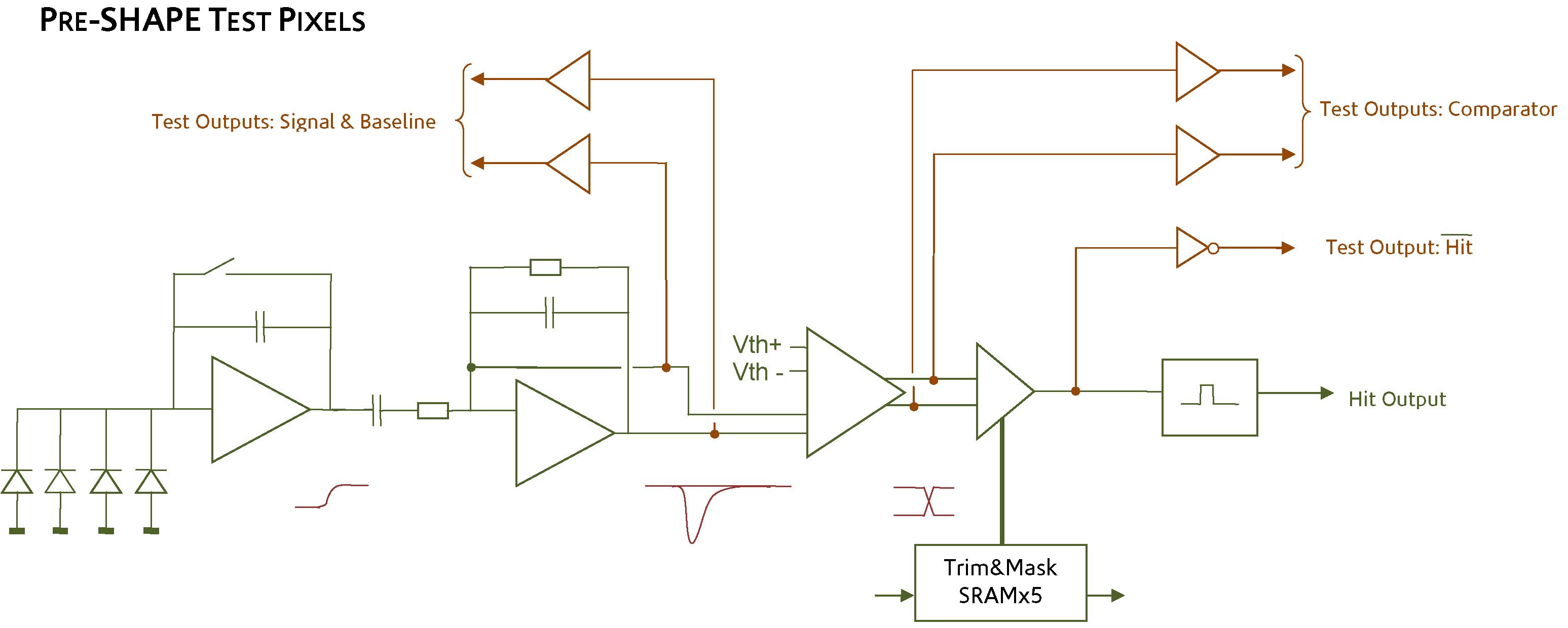}
\includegraphics[width=14cm]{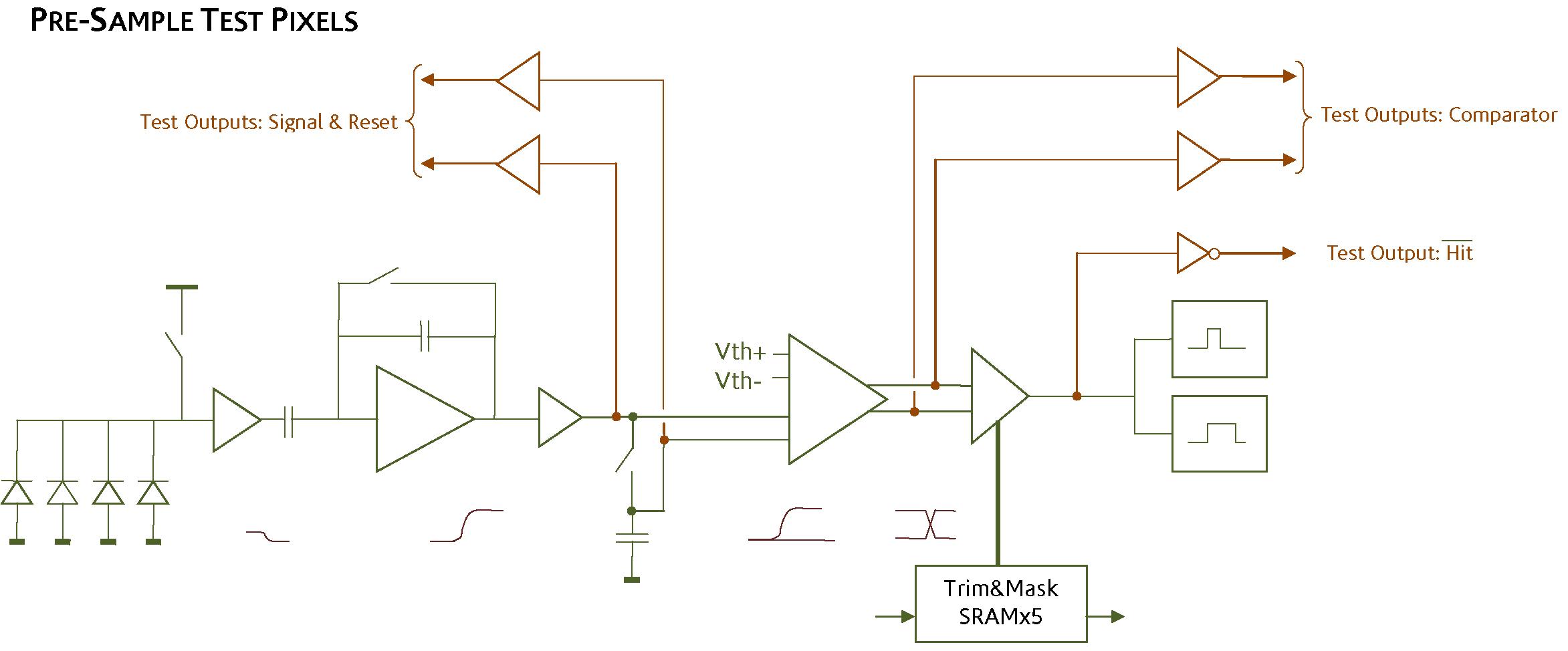}
\caption{\sl 
Test pixel circuits for pre-shape on TPAC~1.1 (upper) and pre-sample 
on TPAC~1.0 (lower), showing the internal nodes
which are accessible externally.
}
\label{fig:design:testpresample}
\end{center}
\end{figure}

In addition to the main design presented above, the TPAC~1.0 sensor has
three pre-sample 
test pixels on the periphery which have been implemented to 
allow access to internal 
nodes for evaluation.
The test pixel designs correspond to both the quadrant~2 and 3 variants 
but also
include additional analogue buffers to monitor internal 
analogue signals in the pixel circuit.
These facilitate evaluation of the 
performance of monostable circuits, comparators, trim adjustment 
of threshold, and the analogue front end circuits for the pre-sample 
pixel architecture. 
The signal pulse and the reset sample are available for two adjacent pixels, 
and the internal differential comparator output is available from one of
these.
A third pixel allows evaluation of other in-pixel circuits, including the 
two monostables and the full comparator chain.
The circuit is shown in
figure~\ref{fig:design:testpresample}.
These test pixels are included at the edge of the main pixel array. Due to
a lack of time before submission, no pre-shaper test pixels were implemented.
However, a pre-shaper test pixel corresponding to the quadrant~1 variant was
included on TPAC~1.1, again with internal signal connections. This is also
shown in figure~\ref{fig:design:testpresample}.

In addition, 
key digital signals, such as control clocks and the least significant bit 
of the multiplex address and time-code are driven off-sensor at the farthest 
point from their initial distribution.
This debug feature allows the timing of critical signals to be evaluated 
during operation.

All bias currents are independent and are generated off-sensor to evaluate the 
performance of sub-circuits in different operating modes.
The two pixel architectures can be operated independently, 
with separate threshold voltage, bias settings and power-down control.

\subsection{DAQ system and operation}
\label{sec:design:daq}

For all the tests described in this paper, the sensor was
mounted on a custom-designed PCB and read out with a
USB-based readout system using a custom software
data acquisition (DAQ) system.

\begin{figure}[ht!]
\begin{center}
\includegraphics[width=8cm]{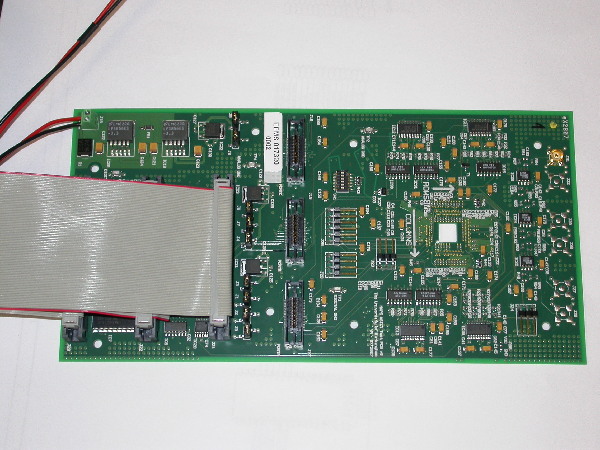}
\caption{\sl 
Photograph of the PCB used to hold and operate the
sensor. The hole with the sensor mounted on the lower
side is just right of centre.
}
\label{fig:design:sensorpcb}
\end{center}
\end{figure}

The PCB design was compatible with both TPAC~1.0 and 1.1 sensors. It
included 31 DACs to allow all the bias
voltages and currents for the sensor to be software
controlled. It also held a temperature sensor. Each
PCB was given a unique ID through a dip switch which
could be read out to the DAQ. Below each sensor, a hole 
approximately $8 \times 8$\,mm$^2$ was cut in the PCB to allow 
access to the substrate for use with a laser 
(see section~\ref{sec:pixel:laser}).
An additional notch approximately $1 \times 1$\,mm$^2$ was 
also cut below the test pixel structures, again to allow illumination
by a laser.
The PCB hole left a strip approximately 1\,mm wide for gluing
the sensor to the PCB. Around 20 sensors were mounted on
PCBs and no mechanical failures of this glue joint were found.
Conductive glue was used to allow the substrate to be
optionally grounded.
The sensor required 
265 wirebonds to connect to the PCB.
The PCB is shown in
figure~\ref{fig:design:sensorpcb}.

Three 64-way flat ribbon cables connected the sensor PCB to a
generic readout board, the USB\_DAQ card, which provided
the USB interface to the computer. The USB\_DAQ card
required firmware specific to the TPAC~1.0 or 1.1 sensor to
generate the required control and timing signals,
as well as the configuration load and readback, and
the data memory readout. 
It also controlled the PCB DAC settings
and performed the temperature and PCB ID readout.

The sensor was operated with fake ``bunch trains'' to mimic
ILC operation. A bunch train consists of a period of data acquisition
where pixel hits are registered and timestamped according to
a ``bunch crossing'' clock.
Following a fixed number of bunch crossings, the data acquisition is
halted and the pixel hits are read out.
The timing structure of the bunch train was controlled
by the USB\_DAQ and was to a large extent under software 
control. For all following measurements, the
operating parameters for the bunch train
used were a 400\,ns period bunch crossing clock 
and a number of bunch crossings up to the
maximum timestamp storable in the memory of 8192.

\section{Pixel charge diffusion simulation}
\label{sec:simulation}


The signal from a minimum ionising particle (MIP)
arises as the particle passes through the sensor.
Physically,
electron-hole pairs are created throughout the silicon
sensor. Within (or near) the epitaxial layer
the liberated electrons are
able to diffuse to the collecting diodes and
hence be seen as signal. 
Some electrons from the 
substrate close to the epitaxial layer can also
diffuse into the epitaxial layer and so be collected.
However, electrons in the deep P-well implant will
mostly be lost.
Overall, for a MIP
at normal incidence to a sensor with
a 12\,$\mu$m
thick epitaxial layer and deep P-well,
around 1100\,$e^-$ will be available
to contribute to the signal. Due to diffusion from
the substrate, a 5\,$\mu$m
thick epitaxial layer will yield around half this value.
Once liberated, the charge diffuses and may be absorbed by 
the collection
diodes of the neighbouring pixel. In addition, other N-well
structures can collect charge and hence charge will be
lost in terms of being signal. This last effect is the
reason for which the deep P-well implant was developed.

As described in Section~\ref{sec:testp:calibration},
for calibration,
the sensor was exposed to an intense $^{55}$Fe source.
This isotope produces gamma radiation including prominent
K$_\alpha$ lines with photon
energies close to 5.9\,keV. There is also a lower rate of
K$_\beta$ photons, with
energies close to 6.5\,keV.
Photons of these energies can interact in
the silicon, again causing ionisation, and will liberate
all the signal charge within a small O(1\,$\mu$m$^3$) volume.
In most cases when the photons convert in the epitaxial layer,
a significant fraction of the charge will diffuse 
out of the pixel and be absorbed in the neighbouring pixels,
being lost as signal to the hit pixel. Hence only a small
fraction of the signal will be collected by the diode and
a spread
of signal sizes will be observed, depending on the details
of the charge diffusion.
However, a
small number of the photons will interact
directly inside the signal diodes themselves.
In this case,
all the charge liberated, around 1620\,$e^-$,
will be collected by the diode and a fixed signal size
will be observed.

\begin{figure}[ht!]
\begin{center}
\includegraphics[width=8cm]{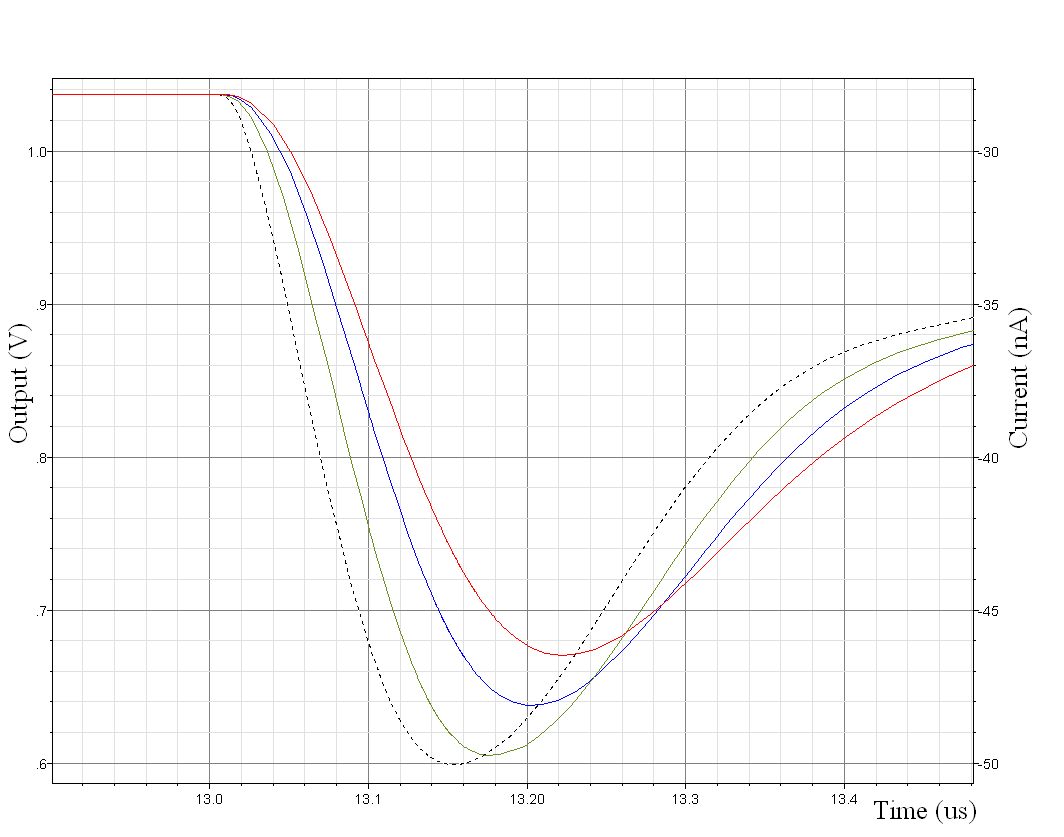}
\caption{\sl 
Simulated reponse of pre-shape circuit to pulses of the same magnitude
but with varying collection times. The pulse charge is injected at 13\,$\mu$s on
the time axis and
collected with an exponential rise, with time constants of 0\,ns (black, dashed),
25\,ns (green), 50\,ns (blue) and 75\,ns (red). Zero corresponds to
a diode interaction while a typical epitaxial interaction corresponds
to the 50 or 75\,ns curve.
}
\label{fig:simulation:time}
\end{center}
\end{figure}

Another difference between an interaction in the diode and elsewhere
in the epitaxial layer is the speed of charge collection.
Charge deposited in the diode should be collected quickly,
of the order of 1\,ns, while the diffusion of charge from the
epitaxial layer takes of order 100\,ns. The diffusion time
is convoluted with the preamplifier response and 
signal shaping circuit, which results in a slower rise time and
lower peak for longer collection times.
Figure~\ref{fig:simulation:time} shows the simulated pre-shaper circuit
response resulting from the different signal time characteristics.
A diode interaction signal is found to be around 15\% larger 
than a more general epitaxial layer interaction, for the same
total charge collected.

\subsection{Detailed simulation}
\label{sec:simulation:sentaurus}

\begin{figure}[ht!]
\begin{center}
\includegraphics[width=7cm]{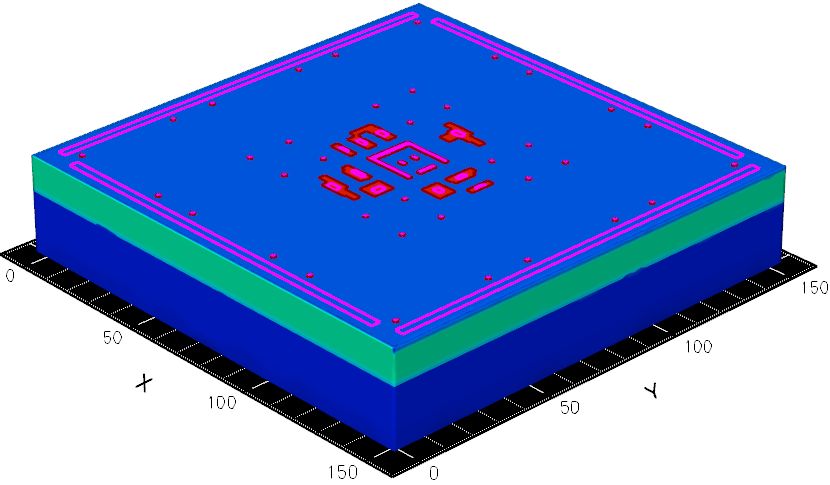}
\includegraphics[width=7cm]{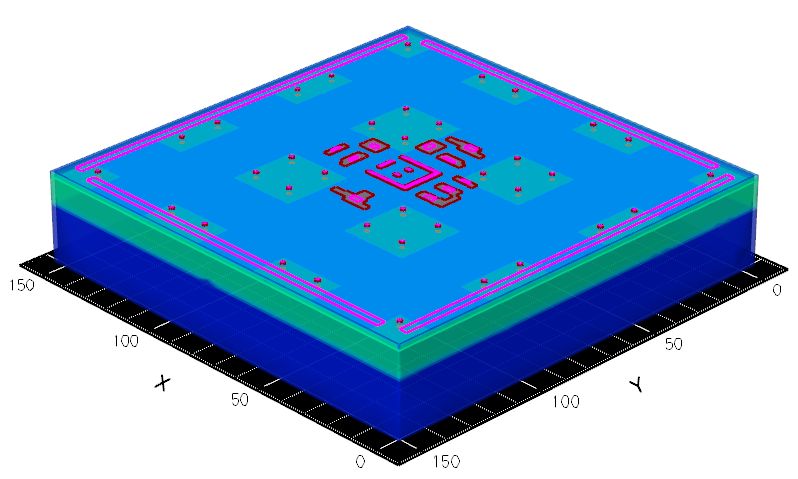}
\caption{\sl 
Diagram of the GDS simulated volume
for the pre-sampler pixel charge diffusion modelling;
without deep P-well (left) and with deep P-well (right).
}
\label{fig:simulation:gds}
\end{center}
\end{figure}

To estimate the signal spread quantitatively, a simulation of
the sensor pixel was performed using the 
Sentaurus~\cite{ref:simulation:sentaurus} package. The pixel
GDS design file was used as input to the simulation, 
ensuring the details of
the pixel were correctly simulated. 
Figure~\ref{fig:simulation:gds} shows the pre-sampler
pixel simulated. The pre-shaper diffusion is expected to
be very similar as the diodes are placed identically in
the two designs; this is particularly true for the deep P-well
sensors as very little charge is expected to be absorbed by
the circuit N-wells. Only modelling of the diffusion of MIP signals
was performed with this detailed simulation.

Charge equivalent to
that expected for a MIP passing perpendicularly through the epitaxial
layer was deposited at various locations
within the pixel and then its diffusion was simulated to investigate where
it was absorbed or lost. A $3\times 3$ array of pixels
was simulated to allow for neighbour pixel charge collection.
The simulation assumed a 15\,$\mu$m epitaxial layer and
simulated the volume
from the sensor surface down to 32\,$\mu$m 
below the surface ($150 \times 150 \times 32$\,$\mu$m$^3$ in total).
A correction is made for all results shown below to account for
the difference between the 15\,$\mu$m epitaxial thickness assumed
in this simulation and the 12\,$\mu$m eventually implemented in the
fabricated sensor.

\begin{figure}[ht!]
\begin{center}
\includegraphics[width=6cm]{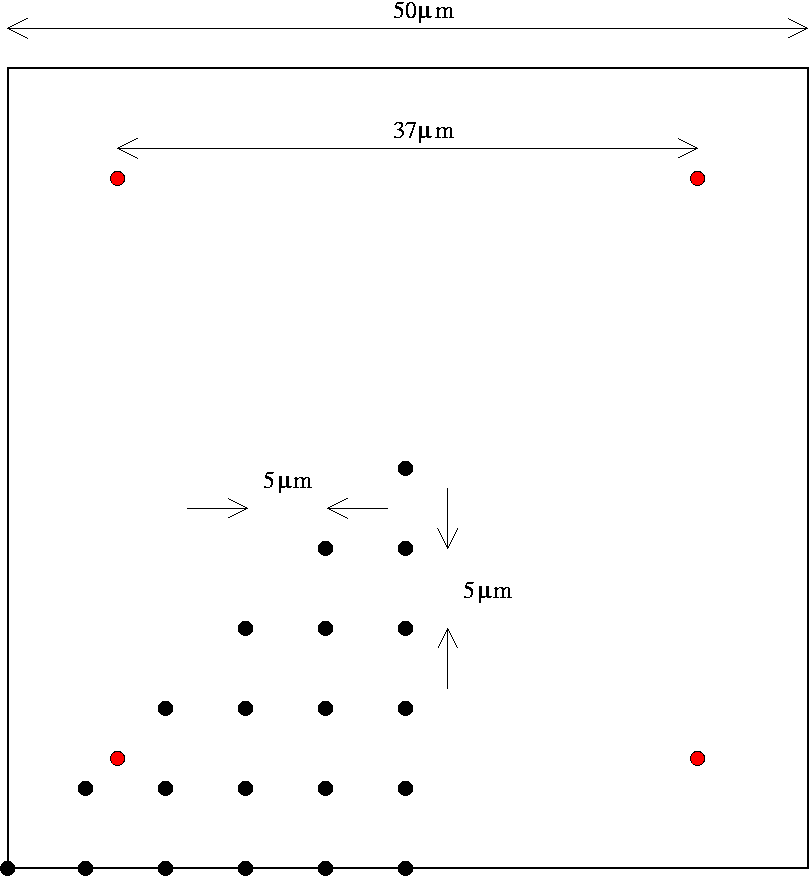}
\caption{\sl 
Layout 
of the 21 points (black circles) used for MIP charge deposits in
the Sentaurus simulation.
The red circles indicate the four diode positions.
}
\label{fig:simulation:points}
\end{center}
\end{figure}

The simulation was CPU-limited so only a small number of MIP
impact positions were modelled. Because of the approximate
eight-fold symmetry of the sensor, all positions
simulated were within a triangle with the pixel centre,
corner, and side centre as its three corners. Within this
triangle, the positions simulated were spaced 5\,$\mu$m
apart, resulting in 21 points in total. Using the 
approximate pixel symmetry,
these could be translated and rotated to give a 5\,$\mu$m
regular array across the whole of the central pixel.
Figure~\ref{fig:simulation:points} shows the layout of the
21 points simulated.

\begin{figure}[ht!]
\begin{center}
\includegraphics[width=7cm]{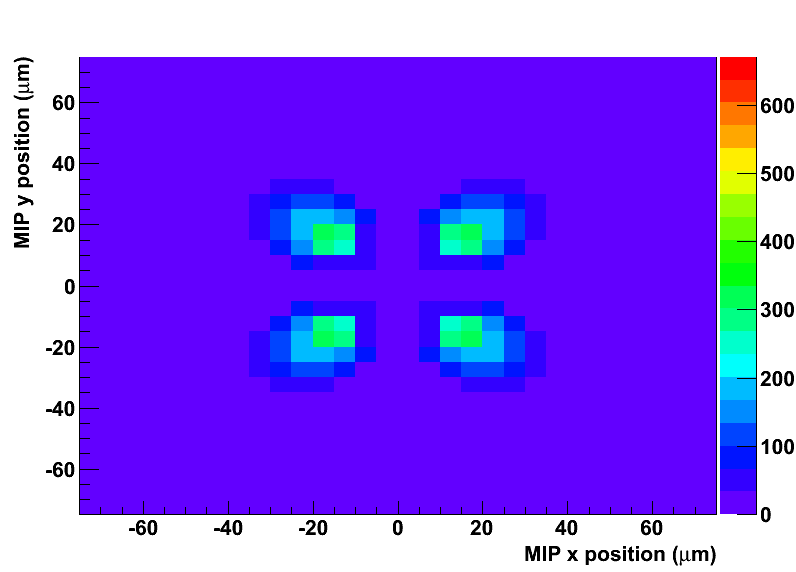}
\includegraphics[width=7cm]{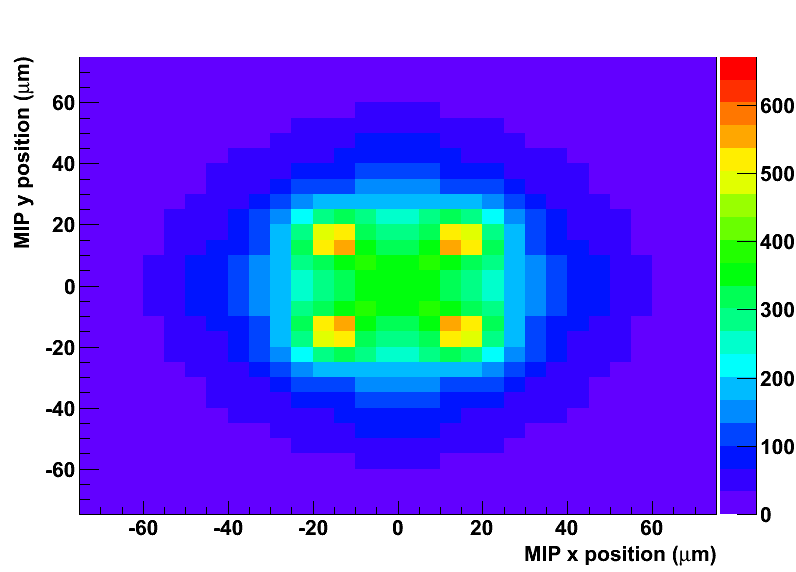}
\caption{\sl 
Sentaurus simulation of 
the MIP charge (in unit of $e^-$) 
seen as signal
for no deep P-well (left)
and deep P-well (right), as a function of the MIP impact position
relative to the centre of the pixel.
The simulation assumed a 15\,$\mu$m thick epitaxial layer, but the
values shown are scaled to those expected for a 12\,$\mu$m layer.
}
\label{fig:simulation:simcharge}
\end{center}
\end{figure}

The simulation
results are shown in figure~\ref{fig:simulation:simcharge}
for sensors with and without the deep P-well implant.
Here the fraction of the signal charge predicted in all nine pixels
of the $3 \times 3$ array
is shown. It is seen that the signal in all cases is predicted to 
be significantly larger with the deep P-well implant. In particular,
in the central pixel, the fraction of charge seen depends on the
position of the incident particle. The signal varies
between 20\% and 50\% with the deep P-well implant, with an average
over the pixel of 30\%. However, without the deep P-well implant,
the central pixel signal can be less than 1\% and only reaches a maximum
of 30\% when near the diode. The average for this case is less
than 10\%.
Similar large differences are seen for the neighbouring
pixels also.

\subsection{Diffusion simulation}
\label{sec:simulation:diffusion}

\begin{figure}[ht!]
\begin{center}
\includegraphics[width=7cm]{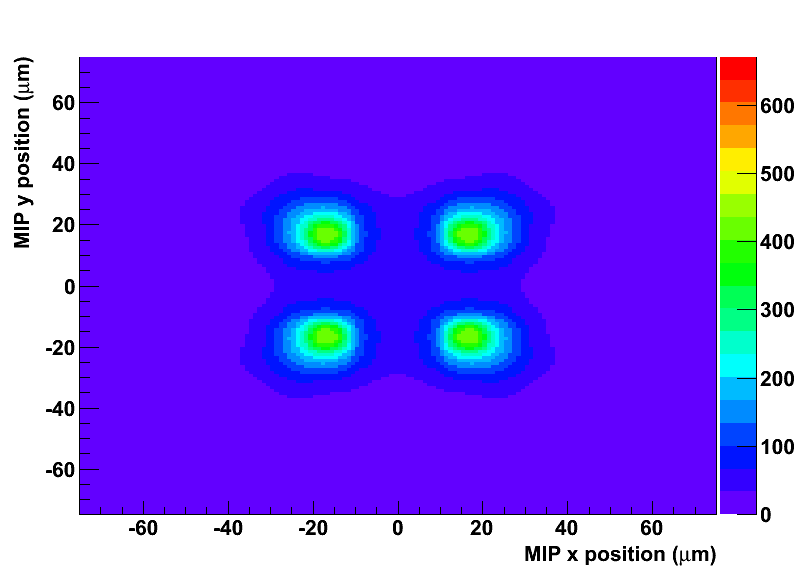}
\includegraphics[width=7cm]{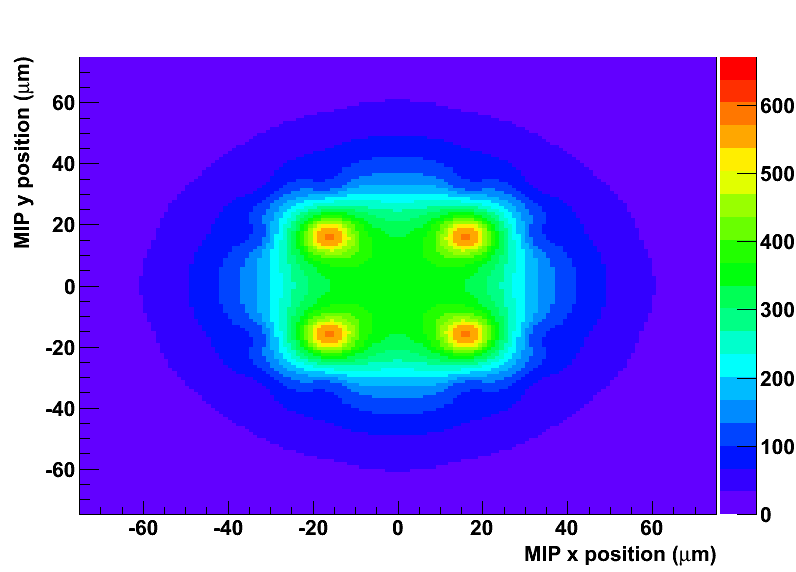}
\caption{\sl 
Simple diffusion simulation of 
the amount of a MIP charge  (in unit of $e^-$) seen as signal
for no deep P-well (left)
and deep P-well (right), as a function of the MIP impact position
relative to the centre of the pixel. The epitaxial layer thickness
is 12\,$\mu$m in these cases.
}
\label{fig:simulation:diffcharge}
\end{center}
\end{figure}

\begin{figure}[ht!]
\begin{center}
\includegraphics[width=7cm]{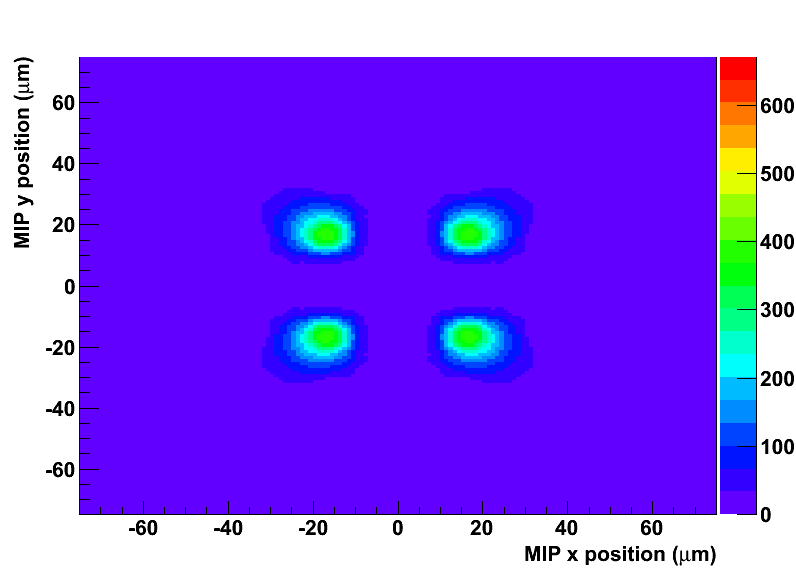}
\includegraphics[width=7cm]{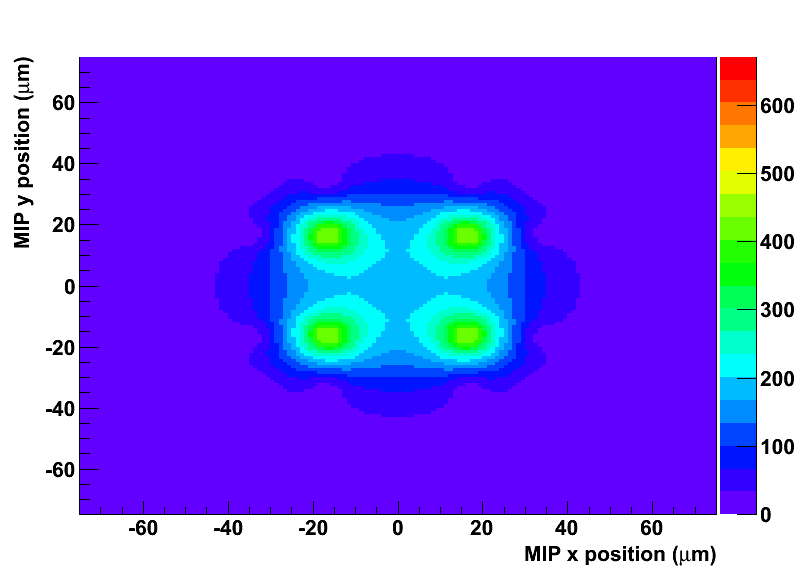}
\caption{\sl 
Simple diffusion simulation of 
the amount of a MIP charge  (in unit of $e^-$) seen as signal
for (left) no deep P-well
and (right) deep P-well, as a function of the MIP impact position
relative to the centre of the pixel. The epitaxial layer thickness
is 5\,$\mu$m in these cases.
}
\label{fig:simulation:diffcharge05}
\end{center}
\end{figure}

The Sentaurus simulation was too time-consuming to model a
large number of charge deposition points,
so a simplified model was used for studying bigger samples of
points and for modelling the $^{55}$Fe calibration response. 
A simple diffusion equation
\begin{displaymath}
\frac{\partial \rho}{\partial t} = - D \nabla^2 \rho
\end{displaymath}
was numerically solved, using an forward-time/centre-difference (FTCD)
method, to model the
charge density $\rho$ within the epitaxial layer.
The diffusion coefficient $D$,
and parameters to model the charge collection with and
without deep P-well implants at the surface, were adjusted
to agree with the results of the Sentaurus simulation for
the above 21 points. The simulation was extended to a
$5 \times 5$ array of pixels and both a 5 or 12\,$\mu$m epitaxial
depth using a 1\,$\mu$m grid. Thus a total of 
$250 \times 250 \times 5$ or $250 \times 250 \times 12$
volume elements were used.
The MIP deposits were again simulated within the same triangle
as shown in figure~\ref{fig:simulation:points} but with 1\,$\mu$m
spacing, giving a total of 351 points. 
Figure~\ref{fig:simulation:diffcharge} shows
the equivalent distributions to figure~\ref{fig:simulation:simcharge}
for MIP deposits resulting from
the simple diffusion model, showing the much finer spacing.
Only the central $3 \times 3$ pixels are shown, although the
simulation volume extended beyond this region.
Figure~\ref{fig:simulation:diffcharge05} shows the same but for
the 5\,$\mu$m epitaxial layer thickness.

\begin{figure}[ht!]
\begin{center}
\includegraphics[width=4.5cm]{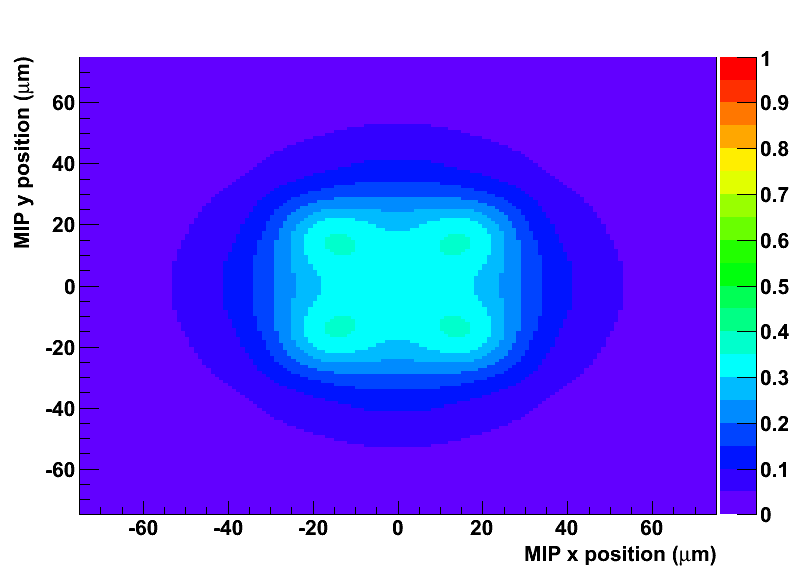}
\includegraphics[width=4.5cm]{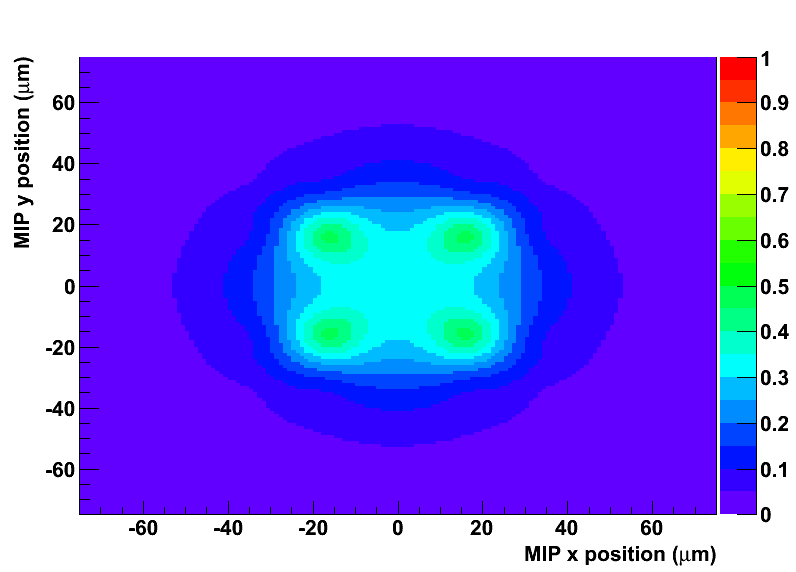}
\includegraphics[width=4.5cm]{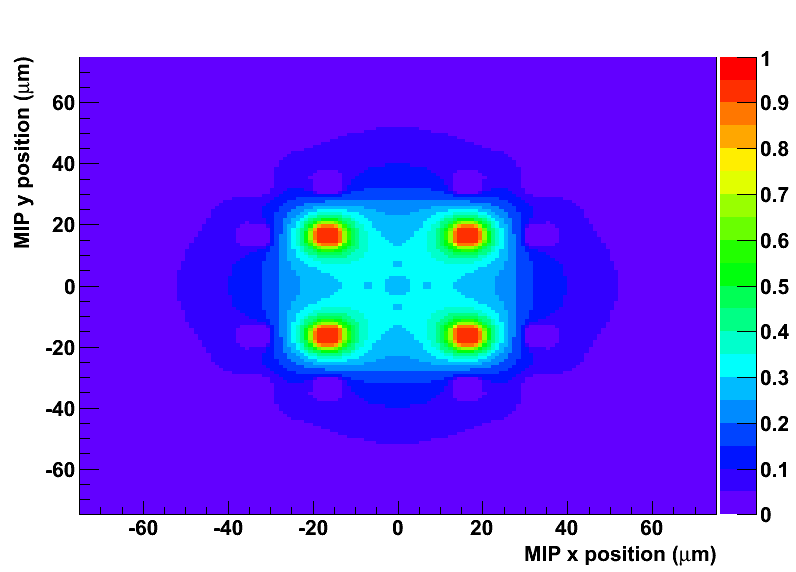}
\caption{\sl 
Simple diffusion simulation of the fraction of the charge seen as signal
for a deep P-well pixel, as a function of the $^{55}$Fe photon impact position
relative to the centre of the pixel at three different depths in the
epitaxial layer. Left: furthest from the diodes, centre: in the middle, right: nearest to the diodes.
The epitaxial layer thickness is 12\,$\mu$m in these cases.
}
\label{fig:simulation:diffdepth}
\end{center}
\end{figure}

For modelling the response to the $^{55}$Fe calibration photons,
the charge deposits
need to be simulated not only over the area of the pixel but also
at various depths in the epitaxial layer.
The simple diffusion model was
used to estimate the spectrum of the charge collected for
randomly positioned charge deposits (as expected for $^{55}$Fe)
throughout the volume of the epitaxial layer. Examples
to illustrate the depth dependence are shown in
figure~\ref{fig:simulation:diffdepth}.

\begin{figure}[ht!]
\begin{center}
\includegraphics[width=4.5cm]{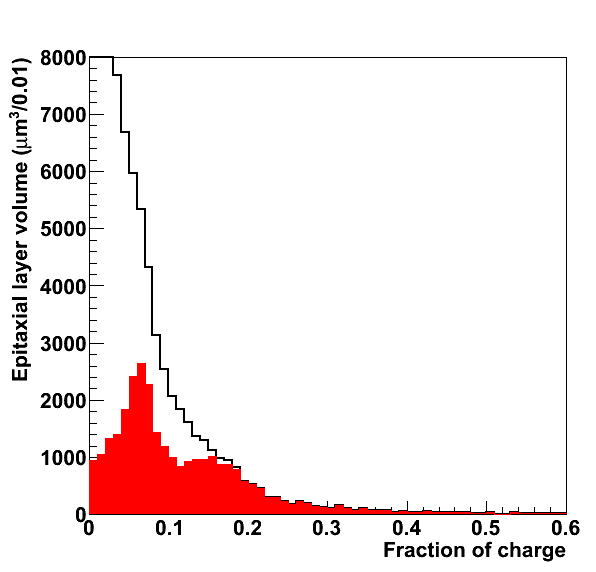}
\includegraphics[width=4.5cm]{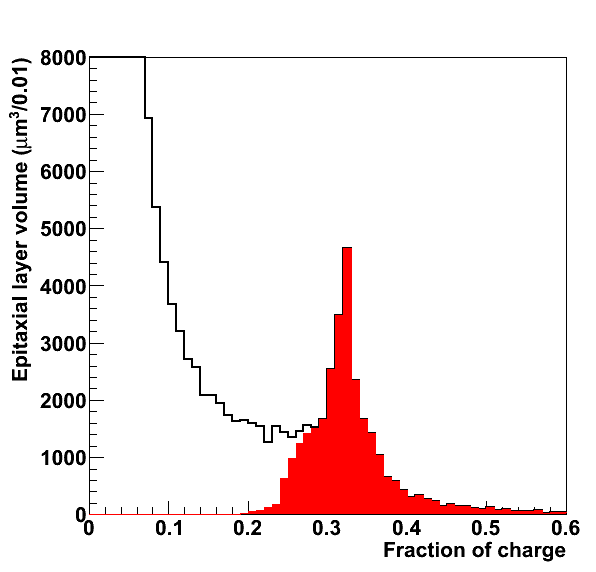}
\includegraphics[width=4.5cm]{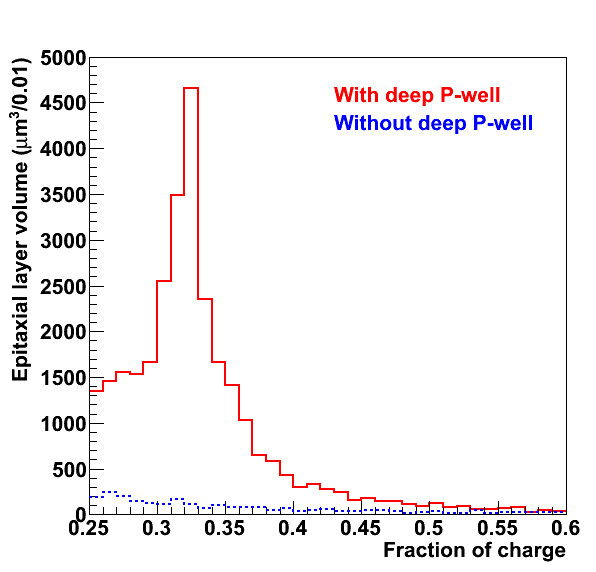}
\caption{\sl 
Simple diffusion simulation of the fraction of the $^{55}$Fe
charge seen as signal
for a uniform distribution of photon hit positions 
throughout the epitaxial volume for no deep P-well (left) and
deep P-well (centre).
The $y$-axis represents the epitaxial layer interaction volume 
of a pixel which results in the given signal fraction.
The filled red histogram shows the fraction of
signal charge absorbed in the central pixel, while the white histogram
shows the spectrum of charge absorbed by the centre pixel as well as the
eight neighbouring pixels. Right: overlaid comparison of the total
distributions from the left and centre plots for
simulated deep P-well (solid, red)
and non-deep P-well (dashed, blue) spectra for an equal irradiation exposure.
The epitaxial layer thickness is 12\,$\mu$m in these cases.
}
\label{fig:simulation:diffspectrum}
\end{center}
\end{figure}

\begin{figure}[ht!]
\begin{center}
\includegraphics[width=4.5cm]{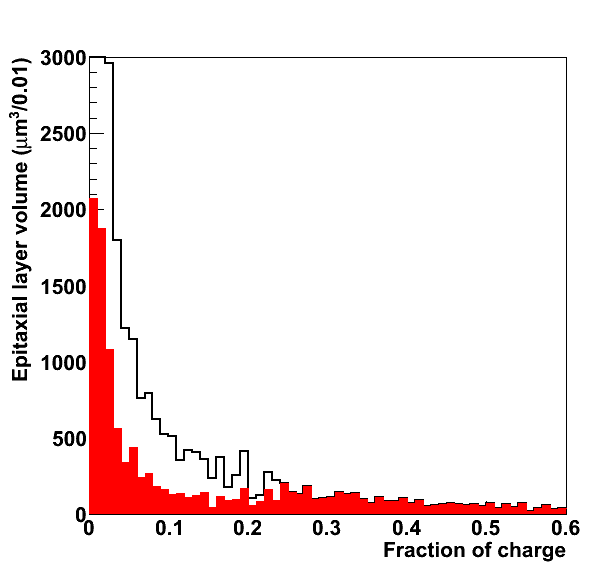}
\includegraphics[width=4.5cm]{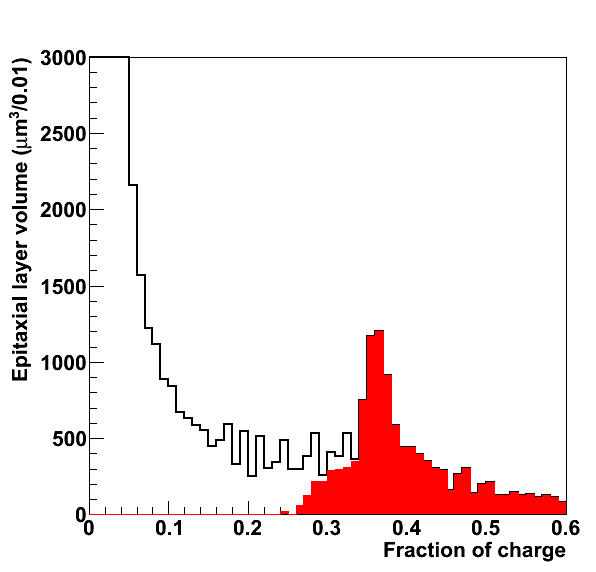}
\includegraphics[width=4.5cm]{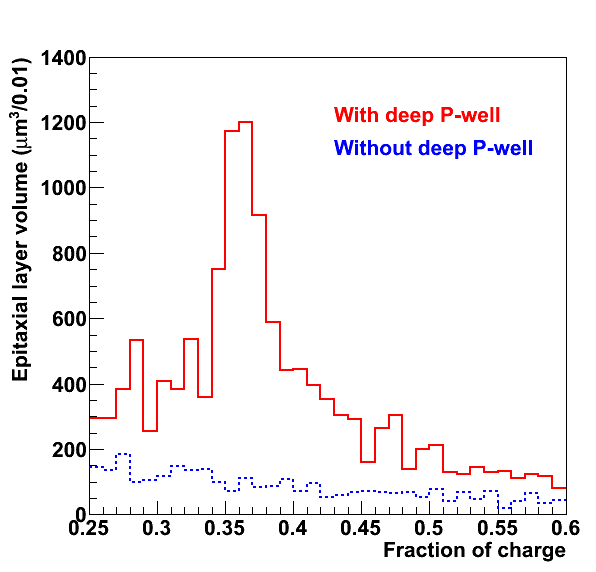}
\caption{\sl 
Simple diffusion simulation of the fraction of the $^{55}$Fe
charge seen as signal
for a uniform distribution of photon hit positions 
throughout the epitaxial volume for no deep P-well (left)
and deep P-well (centre).
The $y$-axis represents the epitaxial layer interaction volume 
of a pixel which results in the given signal fraction.
The red histogram shows the fraction of
signal charge absorbed in the central pixel, while the white histogram
shows the spectrum of charge absorbed by the centre pixel as well as the
eight neighbouring pixels.
Right: overlaid comparison of the total
distributions from the left and centre plots for
simulated deep P-well (solid, red)
and non-deep P-well (dashed, blue) spectra for an equal irradiation exposure.
The epitaxial layer thickness
is 5\,$\mu$m in these cases.
}
\label{fig:simulation:diffspectrum05}
\end{center}
\end{figure}

As can be seen in figure~\ref{fig:simulation:diffdepth},
the simulation
indicates that for the deep P-well cases,
there is a broad ``plateau'' near the centre of
the pixel, effectively independent of epitaxial layer depth, 
where an approximately
constant fraction of the deposited charge is collected. This
spectrum of observed charge is shown in 
figures~\ref{fig:simulation:diffspectrum}
and \ref{fig:simulation:diffspectrum05},
which show the constant fraction is between 0.3 and 0.4,
meaning that a peak corresponding to around 
30-40\% of the $^{55}$Fe energy would be observed.
This fraction is only weakly dependent on epitaxial layer
depth.
(Note, the full energy peak resulting from hits within the signal
diodes is not due to diffusion and so is not seen with this
simple model.) For pixels with no deep P-well, then no such 
plateau peak
is seen in the simulation, so it is expected that only the
full energy peak should be visible in this case. 
Also, for no deep P-well,
the absolute rate of photons with a significant fraction of the full
energy is expected to be significantly lower.

The volume for photon interactions
which give the plateau constant value for the deep P-well case
is a significant fraction
of the epitaxial volume.
For a 12\,$\mu$m epitaxial layer thickness,
it is estimated from the simulation
to be around 30\% of the $50 \times 50 \times 12$\,$\mu$m$^3$
pixel volume, i.e. around $10,000$\,$\mu$m$^3$. 
For the 5\,$\mu$m epitaxial layer thickness case, it is around 25\%,
i.e. $3000$\,$\mu$m$^3$. 
The volume within the diodes
which would give a signal corresponding to the total $^{55}$Fe 
energy is estimated to be approximately a circle of diameter
5\,$\mu$m and depth 1\,$\mu$m, giving a total volume for the
four diodes of around 100\,$\mu$m$^3$. This is independent of
epitaxial layer thickness.
Hence, the plateau region is expected to be
much larger, by around 
two orders of magnitude, than the diodes for 
a 12\,$\mu$m epitaxial layer depth and somewhat smaller for
a 5\,$\mu$m epitaxial layer depth.
Therefore, two peaks are expected from the  $^{55}$Fe
exposure; a large one around 0.3 of the photon energy and 
a much smaller one at the
full photon energy.

\section{Test pixel performance}
\label{sec:testp}

\subsection{Test pixel performance and calibration}
\label{sec:testp:calibration}

As described in section~\ref{sec:design:test},
the sensors included test pixels outside the
$168\times 168$ ``bulk'' pixel array. 
Several internal nodes
of these pixels are accessible externally,
allowing analogue measurements of their levels.

\begin{figure}[ht!]
\begin{center}
\includegraphics[width=8cm]{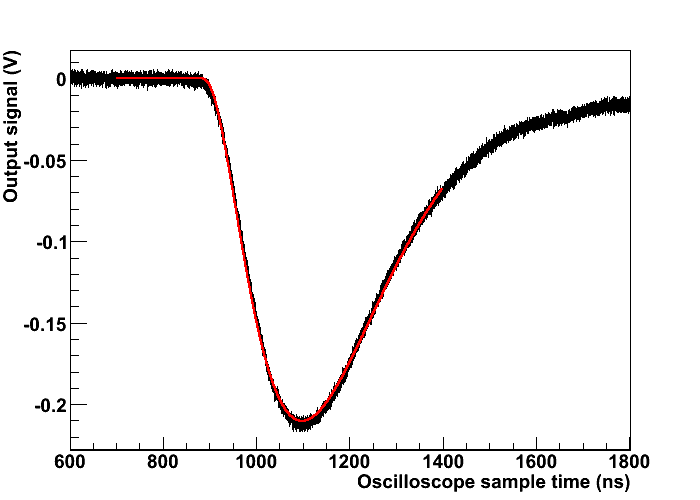}
\caption{\sl 
Typical
deep P-well test pixel oscilloscope signal 
when exposed to $^{55}$Fe photons.
The solid red line shows the result of the
fit performed to extract the peak and
time constants of the signal.
}
\label{fig:testp:scope}
\end{center}
\end{figure}

A low noise pre-amplifier was connected to the output
node from the shaper, the ``Test Output
Signal'' shown in figure~\ref{fig:design:testpresample}
for the pre-shaper on TPAC~1.1,
and the signals were measured on a digital oscilloscope
with a 5\,GHz sampling rate.
%
This system was used to measure the peak signal seen
when the test pixels were exposed to the $^{55}$Fe source.
A powerful GBq
source allowed a sufficient rate of source hits
that a calibration signal could be seen in the
individual test pixels. Two different oscilloscope trigger
threshold levels, 30 and 120\,mV,
allowed the plateau and full energy peaks, respectively, to be measured
separately. 
An exponential rise was convoluted with the pixel circuit
response (the pre-amplifier and CR-RC shaper) and the resulting
functional form fitted to the oscilloscope data to extract the
peak magnitude, the exponential rise time and the start time of the
signal for each trigger.
A typical oscilloscope output and the resulting fit
are shown in figure~\ref{fig:testp:scope}. The shape is seen to have
similar characteristics in terms of the peaking time to the simulated
response shown in figure~\ref{fig:design:preshapemips}.

\begin{figure}[ht!]
\begin{center}
\includegraphics[width=7cm]{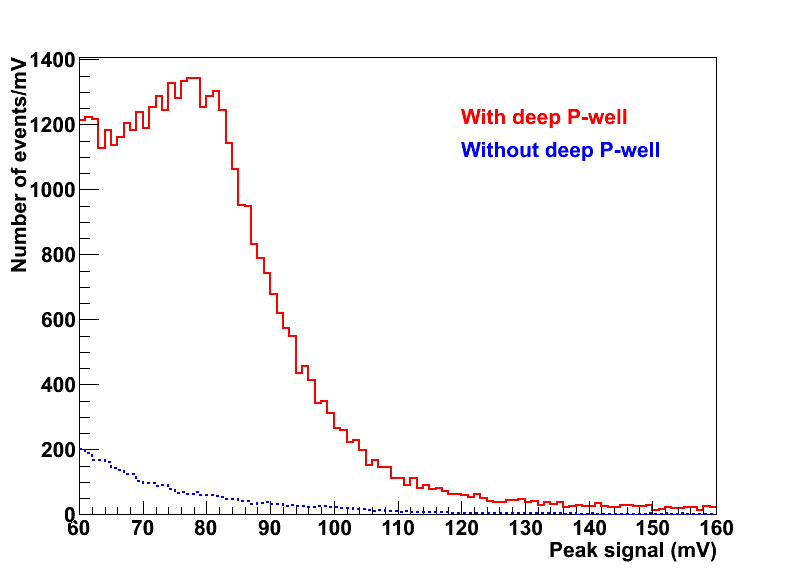}
\includegraphics[width=7cm]{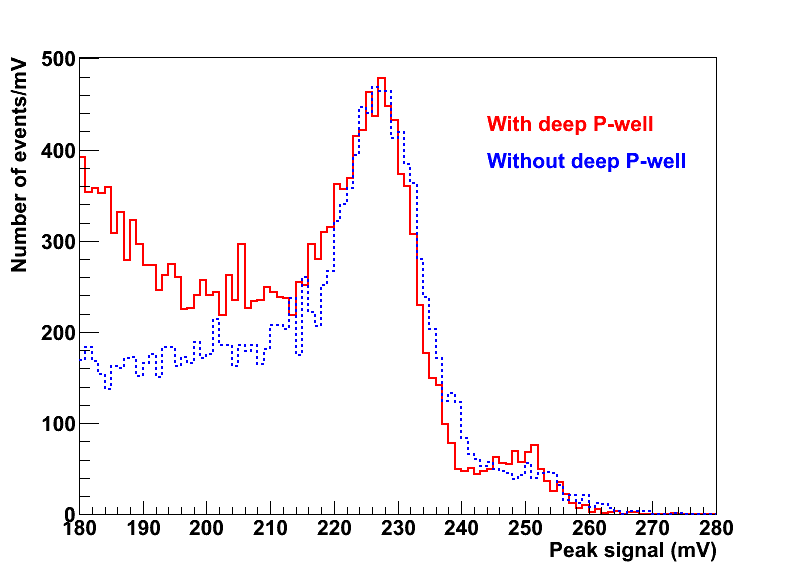}
\caption{\sl 
Test pixel peak signal spectra when exposed to $^{55}$Fe photons
for deep P-well (solid, red) and non-deep P-well (dashed, blue) sensors.
Left: peaks of $^{55}$Fe signal around plateau energy, normalised to
equivalent exposures.
Right: peaks of $^{55}$Fe signal around full energy;
the prominent peaks around 225\,mV are from K$_\alpha$ photons 
of energy 5.9\,keV, while the 
smaller peaks around 250\,mV are due to K$_\beta$ photons of energy 6.5\,keV.
These spectra are from sensors with a 12\,$\mu$m thick epitaxial layer.
}
\label{fig:testp:testiron}
\end{center}
\end{figure}

\begin{figure}[ht!]
\begin{center}
\includegraphics[width=7cm]{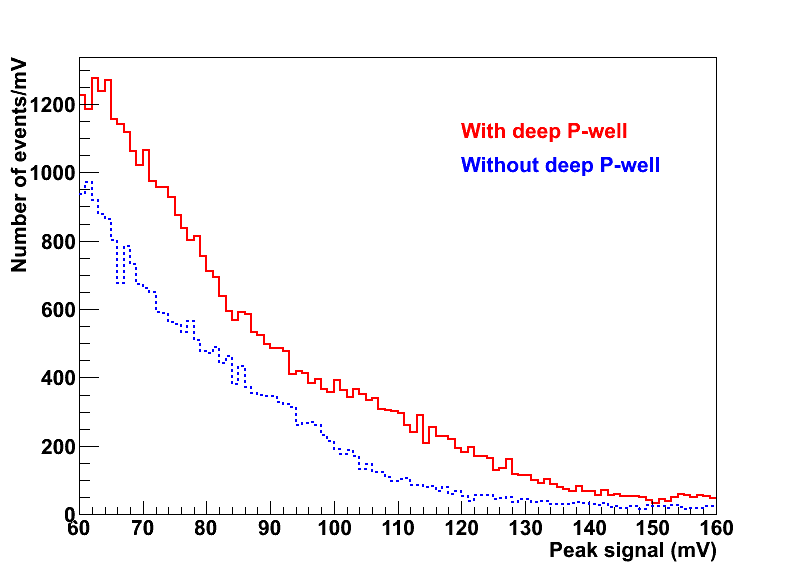}
\includegraphics[width=7cm]{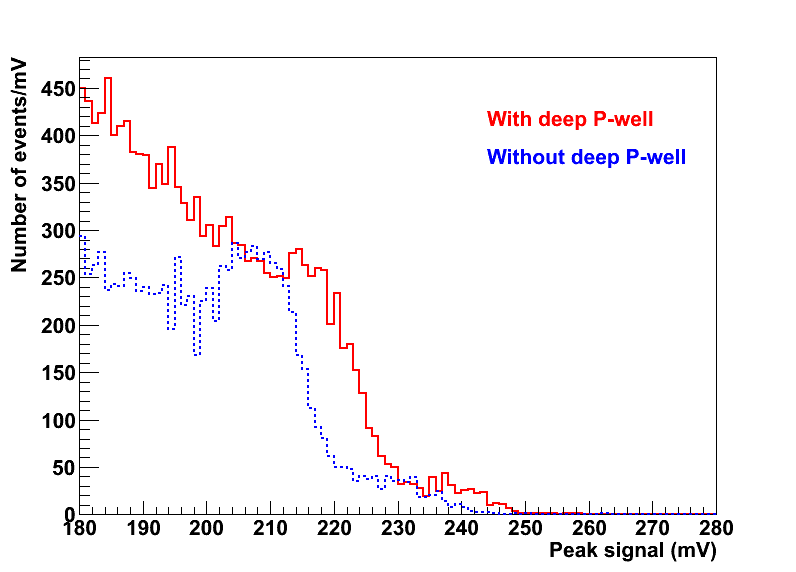}
\caption{\sl 
Test pixel peak signal spectra when exposed to $^{55}$Fe photons
for deep P-well (solid, red) and non-deep P-well (dashed, blue) sensors.
Left: peaks of $^{55}$Fe signal around plateau energy, normalised to
equivalent exposures. 
Right: peaks of $^{55}$Fe signal around full energy;
the peaks around 215\,mV are from K$_\alpha$ photons 
of energy 5.9\,keV, while the
smaller peaks around 235\,mV are due to K$_\beta$ photons of energy 6.5\,keV.
These spectra are from sensors with a 5\,$\mu$m thick epitaxial layer.
}
\label{fig:testp:testiron05}
\end{center}
\end{figure}

The spectra of the magnitude of the fitted peak signals observed 
in 12\,$\mu$m epitaxial layer sensors
are shown in figures~\ref{fig:testp:testiron}
and \ref{fig:testp:testiron05}.
The full energy peaks are similar with or without deep P-well
as these mainly result from photon interactions in the diode,
where there is no deep P-well implant.
Comparing the plateau peaks to the simulation,
the data spectra correspond roughly to an equivalent charge fraction
as shown in figure~\ref{fig:simulation:diffspectrum},
and \ref{fig:simulation:diffspectrum05}.
For the 12\,$\mu$m epitaxial thickness case,
it is seen that the plateau peak position
and the relative rates for deep P-well and no deep P-well are
in reasonable agreement, although the plateau peak width is clearly
much broader in the data. Given the simple level of
simulation used, this is not unreasonable.
For the 5\,$\mu$m epitaxial thickness case, there is no clear
peak; the absence of the peak predicted by the simulation is 
not understood.

The noise measured by the oscilloscope is 3\,mV. Both this noise
and the
full energy peak positions are consistent both with and without
deep P-well , showing the deep P-well implant
does not affect the circuit performance. 
For the deep P-well,  $12\,\mu$m epitaxial
thickness, the full peak corresponds to
$1620\,e^-$ charge, so the noise is equivalent to $22\,e^-$.
Allowing for a gain of 0.8 in the external low-noise preamplifier,
the pre-shaper gain is measured to be $160\,\mu$V$/e^-$.
These values are
completely consistent with the expectations given in
section~\ref{sec:design:preshape}.
In addition, the ratio
of the plateau to full peaks is 34\%
which is in
reasonable agreement with the simple diffusion simulation,
see figure~\ref{fig:simulation:diffspectrum}. 
Given the
approximations used in this model, the agreement is regarded
as sufficient to be confident that the source of these
peaks is understood.

\begin{figure}[ht!]
\begin{center}
\includegraphics[width=7cm]{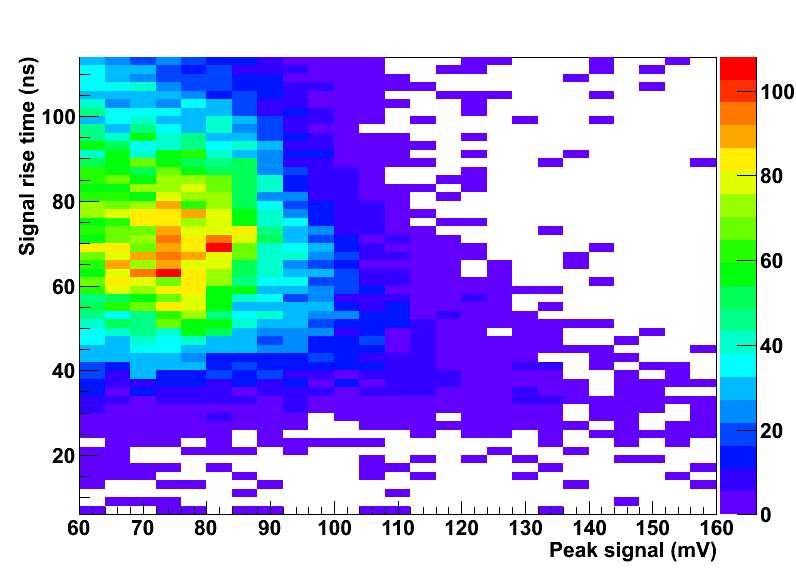}
\includegraphics[width=7cm]{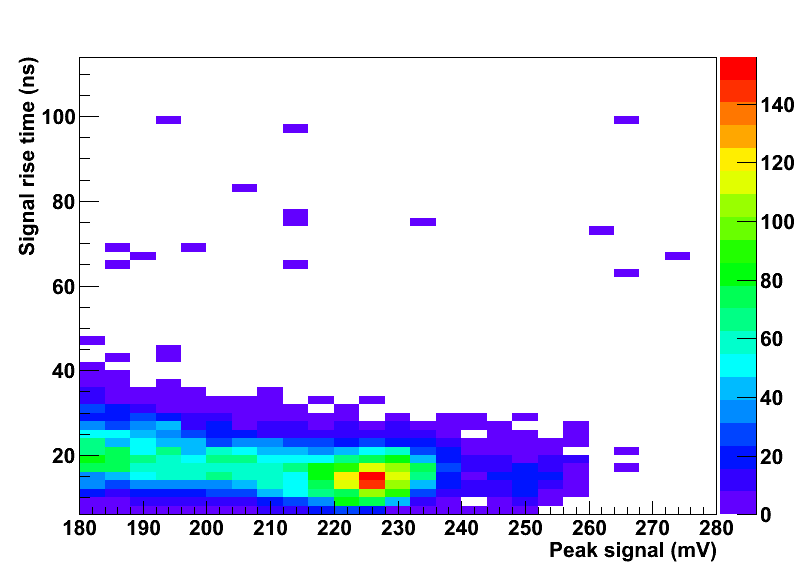}
\caption{\sl 
Deep P-well test pixel signals when exposed to $^{55}$Fe photons,
showing the number of events (indicated by the colour scale)
seen as a function of the signal
peak value and exponential time constant, both extracted from
a fit. Left: the plateau region. Right: the full peak region.
These data are from a sensor with a 12\,$\mu$m deep epitaxial layer.
}
\label{fig:testp:irontime}
\end{center}
\end{figure}

As a final check of the origin of these peaks, 
figure~\ref{fig:testp:irontime}
shows the correlation of the fitted exponential
rise time constants of the signals as a function of the
fitted peak signal magnitude. 
The full peak values have rise times of around 10\,ns
while the deposits resulting in the plateau peak have a rise time
of around 70\,ns.
The strong correlation shows the highest peak values, which correspond to
the diode interactions, have the shortest time constant, as expected,
while the plateau signal, from lower in the epitaxial layer, has a slow
collection time;
see section~\ref{sec:simulation}.

\section{Bulk single pixel performance}
\label{sec:pixel}

\subsection{Pedestals and noise}
\label{sec:pixel:peds}

The performance in the bulk pixels is less straightforward
to measure as there is no analogue readout available. However,
the rate of hits in each pixel as a function of the applied
threshold effectively allows the integral of the analogue
spectrum to be measured and hence the spectrum itself can
be estimated. The results in this section are based on 
performing such ``threshold scans'' and measuring the response.

One further complication is that pickup was observed between pixels
when O(100) were enabled and firing at the same time (see
section~\ref{sec:sensor:pickup}). This is not a major problem when
operating the sensor for its design purpose as only O(10)
pixels per sensor
are expected to fire in each event. However, for the basic
performance measurements presented below, this pickup could
prevent the response of individual pixels from being measured.
Hence to remove this sensor-wide effect, only 168
pixels were unmasked in any run for the following results, and
hence 168 such runs were required for a complete threshold scan of 
the whole sensor. As shown in section~\ref{sec:sensor:pickup}, this
number is sufficiently small that pickup effects do not occur.

To select the 168 unmasked pixels, only one pixel of the 42 which 
shared SRAM memory was unmasked in each run. In addition, runs were 
performed with only 19 bunch crossings per bunch train. Both of
these together ensured that the memory could never overflow
and so distort the threshold scan, even when very close to
threshold.

\begin{figure}[ht!]
\begin{center}
\includegraphics[width=6cm]{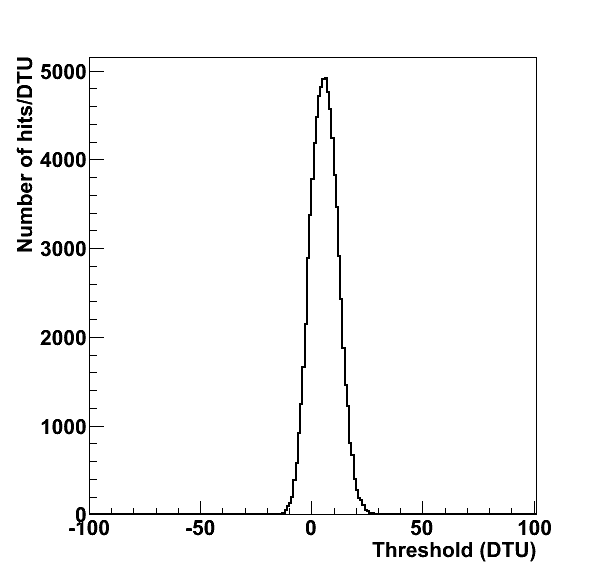}
\caption{\sl 
Typical pixel threshold scan for a pre-shape pixel.
}
\label{fig:pixel:typicaltscan}
\end{center}
\end{figure}

The pixel comparator only fires when the input crosses through
the threshold from below (see section~\ref{sec:design:preshape}).
In contrast to a level comparator,
this means the comparator will not fire if the input is far 
from the threshold with either polarity. Hence,
a threshold scan performed with no external stimulus will
result in hits only when near the pixel pedestal.
The results of such a scan for a typical pixel
are shown in figure~\ref{fig:pixel:typicaltscan}. 
Here, the scan is specified in the (arbitrary) units of the DAC 
threshold setting, called ``DAC threshold units'' (DTU) in this paper.

\begin{figure}[ht!]
\begin{center}
\includegraphics[width=6cm]{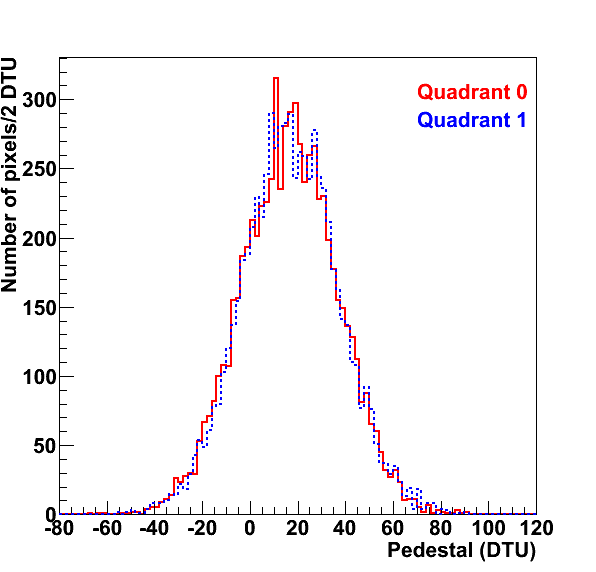}
\includegraphics[width=6cm]{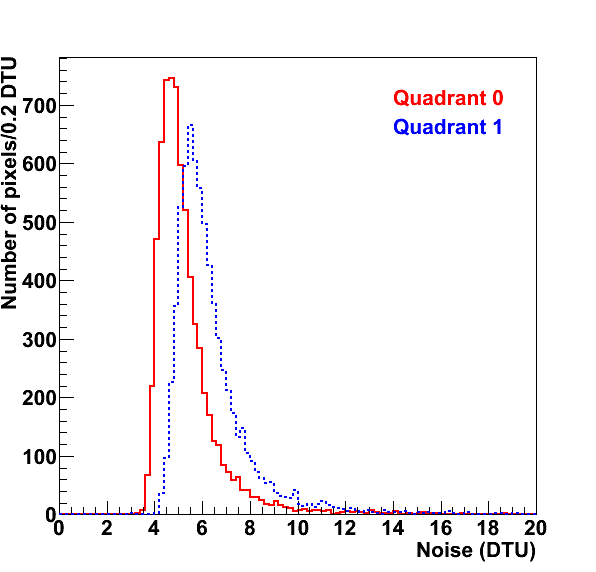}
\caption{\sl 
Distributions of pedestals (left) and noise (right)
for pixels in quadrants~0 (solid, red) and 1 (dashed, blue), in DAC threshold
units (DTU), for a typical sensor. Pixels in these two quadrants are all
of the pre-shape design. All trim values were set to zero.
}
\label{fig:pixel:pedsquad}
\end{center}
\end{figure}

For the pre-shape threshold scan, the
resulting distribution is well described by a Gaussian and
so can be characterised by the mean and the RMS.
The mean gives the pedestal for the pixel and the RMS gives
the noise~\cite{ref:pixel:noise}.
The distribution of the pixel mean and RMS values for
the two pre-shape quadrants (see Section~\ref{sec:design:preshape})
of a typical sensor are
shown in figure~\ref{fig:pixel:pedsquad}.
It is seen that there is a significant spread of pedestals
within a quadrant. The pedestal spread RMS is around 20\,DTU
and is the same for quadrants~0 and 1.
This is
much bigger than the noise on the pixels, which has an average
value of around 5\,DTU for quadrant~0 and 6\,DTU for quadrant~1.

\begin{figure}[ht!]
\begin{center}
\includegraphics[width=6cm]{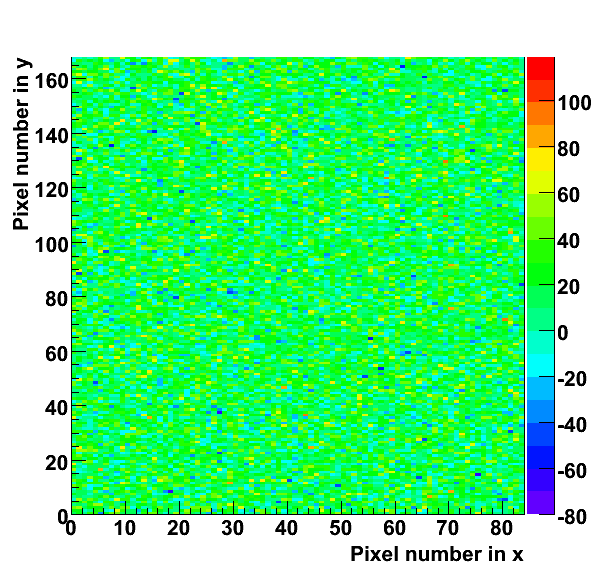}
\includegraphics[width=6cm]{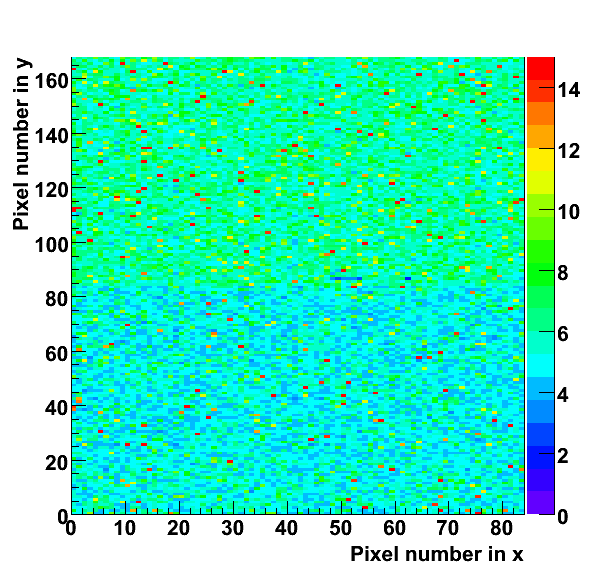}
\caption{\sl 
Two-dimensional map of pedestals (left) and noise (right) for 
the two pre-shape quadrants of a typical sensor.
The lower half of each plot corresponds to quadrant~0
and the upper half to quadrant~1. The colour-coded scale
is in DAC threshold units (DTU) in both cases.
}
\label{fig:pixel:twodpedestals}
\end{center}
\end{figure}

The values of the pedestal and noise for each pre-shape pixel 
of a typical sensor
as a function of the position within the sensor are shown in 
figure~\ref{fig:pixel:twodpedestals}, where the difference
in noise for the two quadrants is visible.
The values of the pedestals and noise show no correlation 
with position within each quadrant and their spreads appear 
genuinely random pixel-to-pixel. There is no correlation
between the pedestal and noise values.

\begin{figure}[ht!]
\begin{center}
\includegraphics[width=6cm]{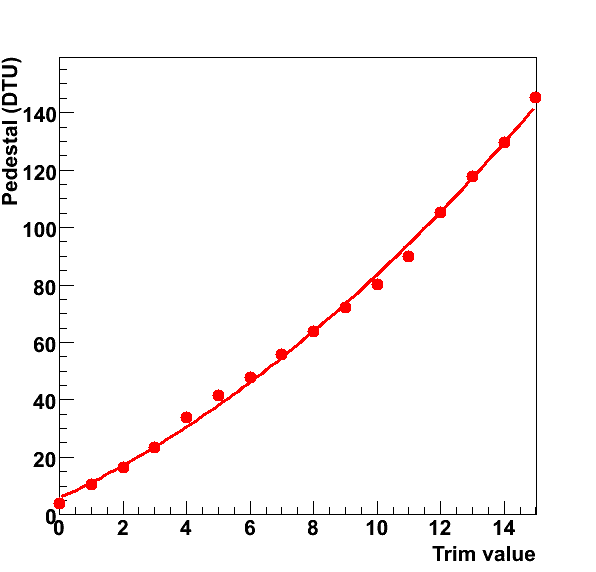}
\caption{\sl 
Change of pedestal for a typical pixel
as a function of the trim setting. The curve shows a
second-order polynomial fit to the data points,
illustrating the response is non-linear.
}
\label{fig:pixel:typicaltrimscan}
\end{center}
\end{figure}

The spread of pedestals can be reduced using the per-pixel
trim setting. This allows a value in the range 0-15 to be
loaded to each pixel to adjust the pedestal position upwards.
The effect of this on the pedestal of a typical pixel is shown
in figure~\ref{fig:pixel:typicaltrimscan}. It is seen that
there is a non-linear effect in the response.

\begin{figure}[ht!]
\begin{center}
\includegraphics[width=6cm]{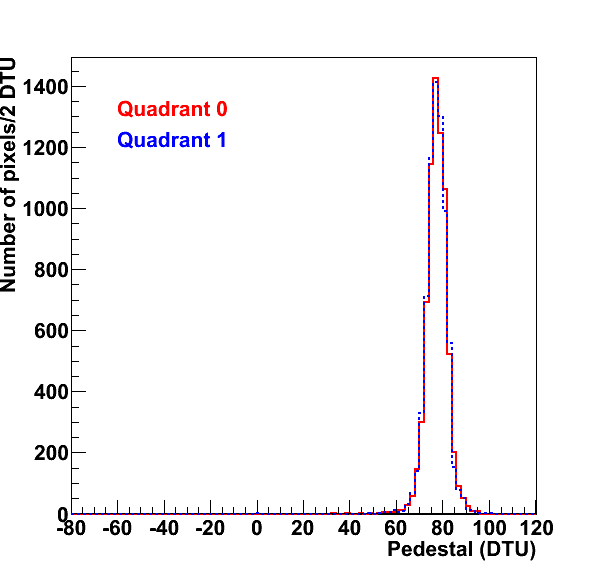}
\includegraphics[width=6cm]{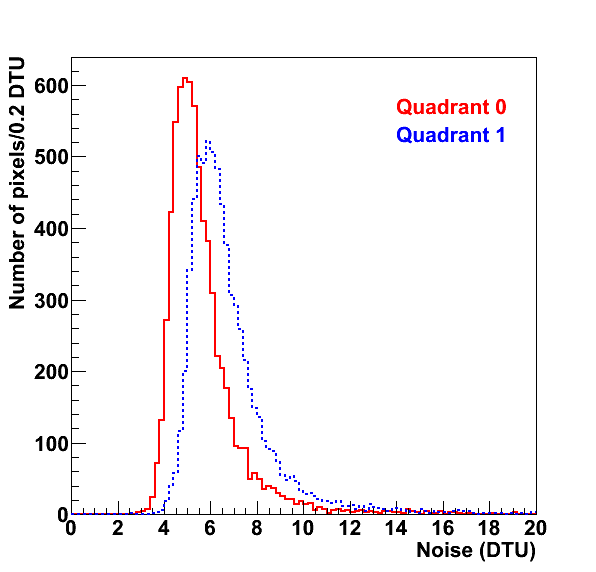}
\caption{\sl 
Distributions of pedestals (left) and noise (right)
for pixels in quadrants~0 (solid, red) and 1 (dashed, blue), in DAC threshold
units (DTU), for a typical sensor, following the determination of the
optimal trim values.
}
\label{fig:pixel:trimadjust}
\end{center}
\end{figure}

The spreads of pedestals after trimming a
typical sensor are shown in figure~\ref{fig:pixel:trimadjust}.
It is seen that the RMS of the pedestals is reduced by
a factor of around five. The resulting spread is a little
smaller than the
size of the per-pixel noise and so gives a small but
non-negligible
contribution to the apparent noise rate when setting a common
threshold for all pixels. The ability to improve on
this effect is limited by the
number of trim bits available. With four bits, then a
granularity of the trim of 16 values is possible. However,
the total range of the uncorrected pedestals is around 100\,DTU, 
so that even with careful matching of the trim range, the trim
least significant bit 
would be around 6\,DTU. Two additional trim bits have been
implemented in the next version of the sensor.
The average noise is unaffected by trimming, although the
spread of the noise is slightly increased.

\begin{figure}[ht!]
\begin{center}
\includegraphics[width=6cm]{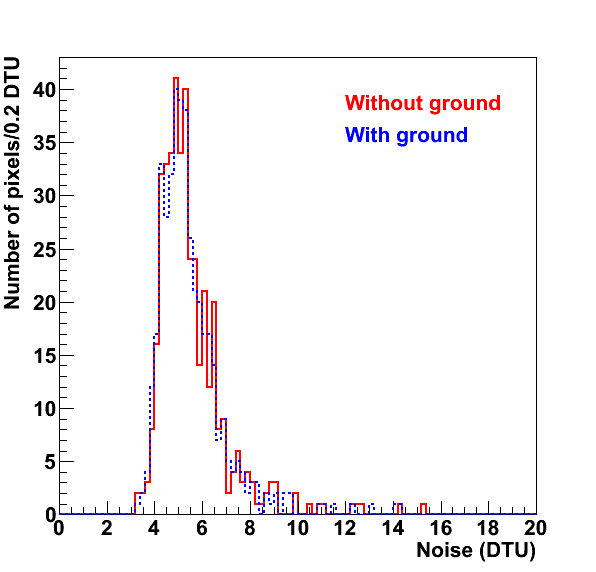}
\includegraphics[width=6cm]{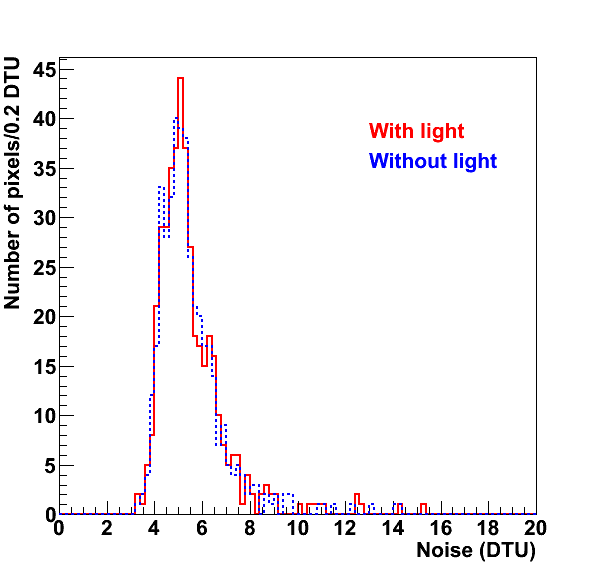}
\caption{\sl 
Left: noise distribution with (dashed, blue) and without (solid, red) grounding the substrate.
Right: noise distribution with (solid, red) and without (dashed, blue)
a strong light source shining on the sensor.
}
\label{fig:pixel:substratenoise}
\end{center}
\end{figure}

We investigated the pixel noise under a variety of external factors. 
The sensor substrate
can be grounded or operated with no explicit bias connection to 
the substrate.
This could in principle change any environmental noise pickup.
The distribution of noise for a subset of pixels both
with and without the substrate being grounded are shown in
figure~\ref{fig:pixel:substratenoise}. It is seen that leaving
the substrate unconnected has no significant effect on the
noise. In addition, the sensor may be sensitive to light
and it is normally operated with a cover to keep it in the
dark. However,
the six metal layers used effectively completely cover
the top surface of the sensor so little light
reaches the epitaxial layer.
Figure~\ref{fig:pixel:substratenoise}
also shows the noise distribution of the same pixels with
and without a strong lamp being shone on the top surface of
the sensor. Again, there is no significant difference.


\subsection{Bulk pixel calibration}
\label{sec:pixel:calib}

\begin{figure}[ht!]
\begin{center}
\includegraphics[width=6cm]{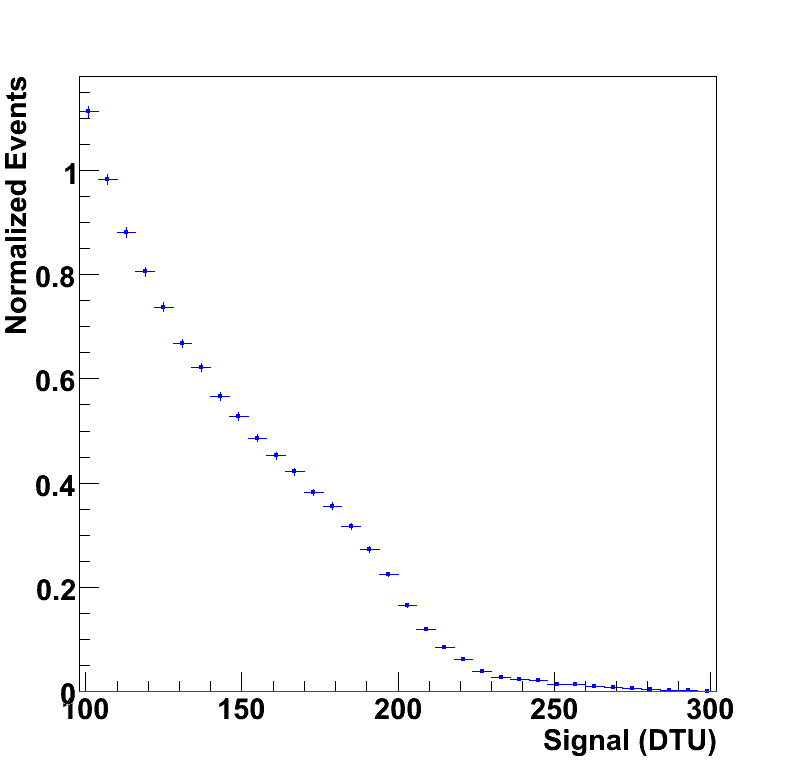}
\includegraphics[width=6cm]{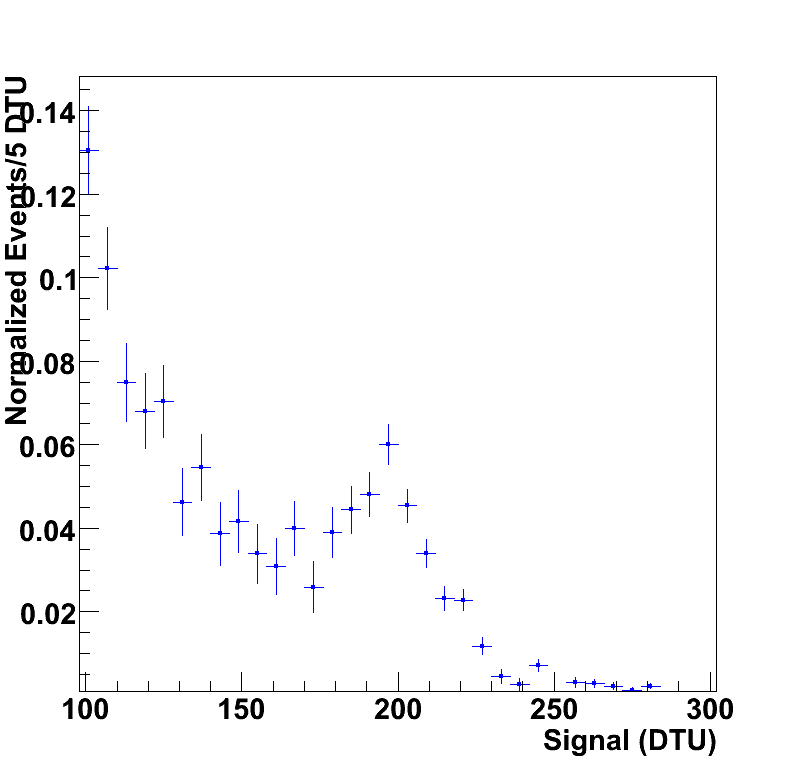}
\caption{\sl 
Left: rate of hits as a function of the threshold setting.
Right: numerical derivative of the rate plot. The $^{55}$Fe
calibration signal is visible at around 190\,DTU, as a shoulder
in the left plot but as a clear peak in the right plot.
The epitaxial layer thickness is 12\,$\mu$m in this case.
}
\label{fig:pixel:iron}
\end{center}
\end{figure}

The calibration of the threshold setting in DTU to a physical
scale was performed using an $^{55}$Fe source using
the same principle as described in section~\ref{sec:testp:calibration}.
The same high-rate source was used, 
giving a sufficient rate of hits to calibrate
individual pixels.
A threshold scan was performed and the rate plotted as
a function of the threshold setting. A numerical derivative
was then taken of this rate plot to get the basic spectrum
and so to find the calibration peak. 
The results for a typical pixel are shown in
figure~\ref{fig:pixel:iron}. The observed peak corresponds to
the plateau peak, as the full energy peak is above the operating
range of the comparator.

\begin{figure}[ht!]
\begin{center}
\includegraphics[width=6cm]{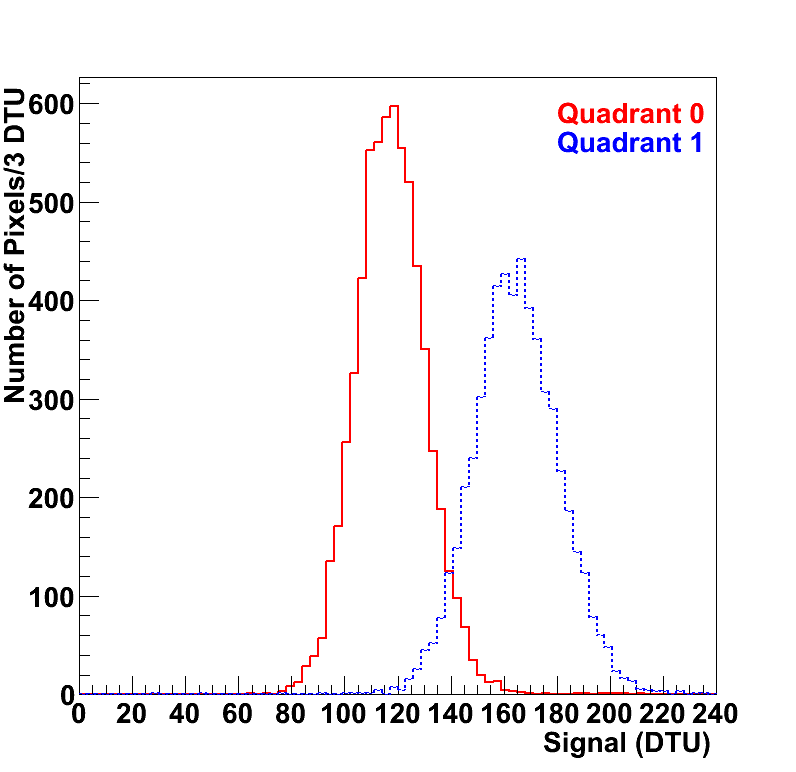}
\includegraphics[width=6cm]{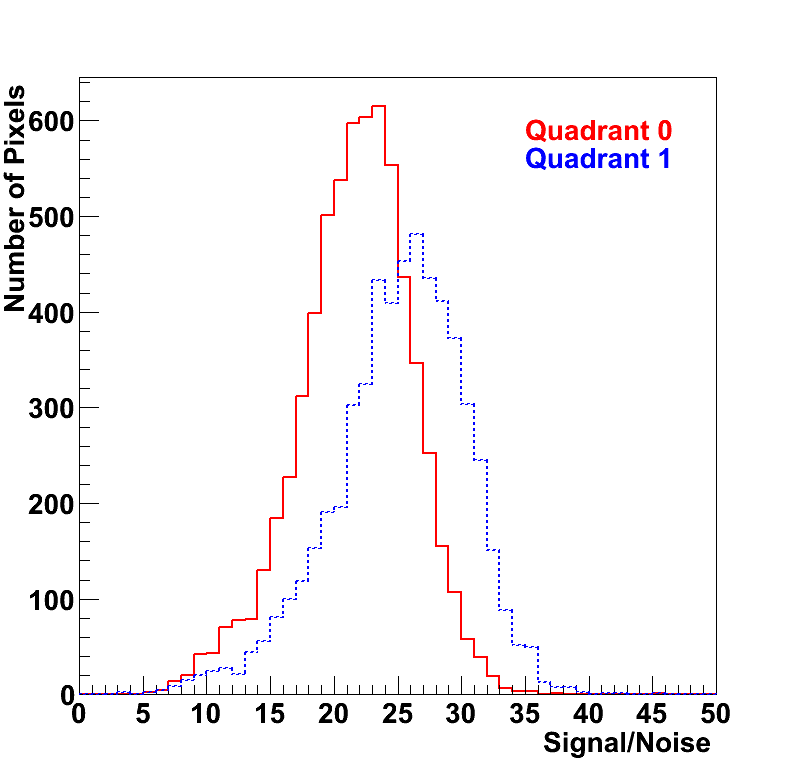}
\caption{\sl 
Distribution of $^{55}$Fe signal peak position (left) 
and signal/noise (right) for pixels in
quadrants~0 (solid, red) and 1 (dashed, blue). 
}
\label{fig:pixel:irongain}
\end{center}
\end{figure}

The peak positions (corrected for pedestal) for the pre-shape pixels 
measured in both
quadrants~0 and 1 are shown in figure~\ref{fig:pixel:irongain}.
The average signal for quadrant~1, 165\,DTU, is significantly higher 
than for quadrant~0, 117\,DTU,
by around 35\%, while
the spreads of the signals for quadrants~0 and 1
are 11\% and 10\%, respectively.
As shown above in figure~\ref{fig:pixel:pedsquad},
quadrant~1 also has a higher average noise than quadrant~0, by around 20\%,
so quadrant~1 has a better signal/noise ratio (averaging 25 compared with 22),
also shown explicitly in
figure~\ref{fig:pixel:irongain}.

\begin{figure}[ht!]
\begin{center}
\includegraphics[width=6cm]{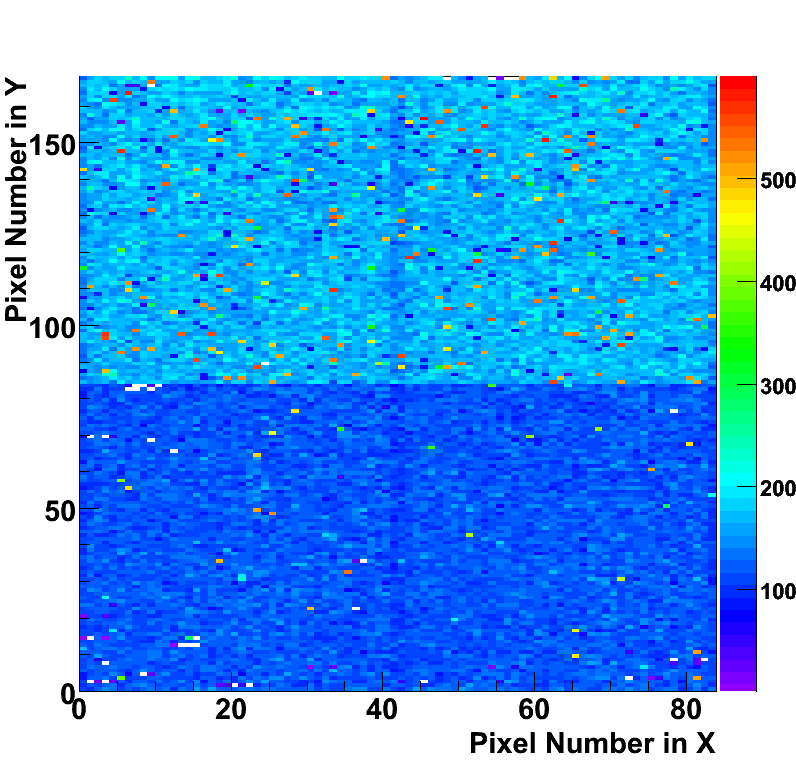}
\includegraphics[width=6cm]{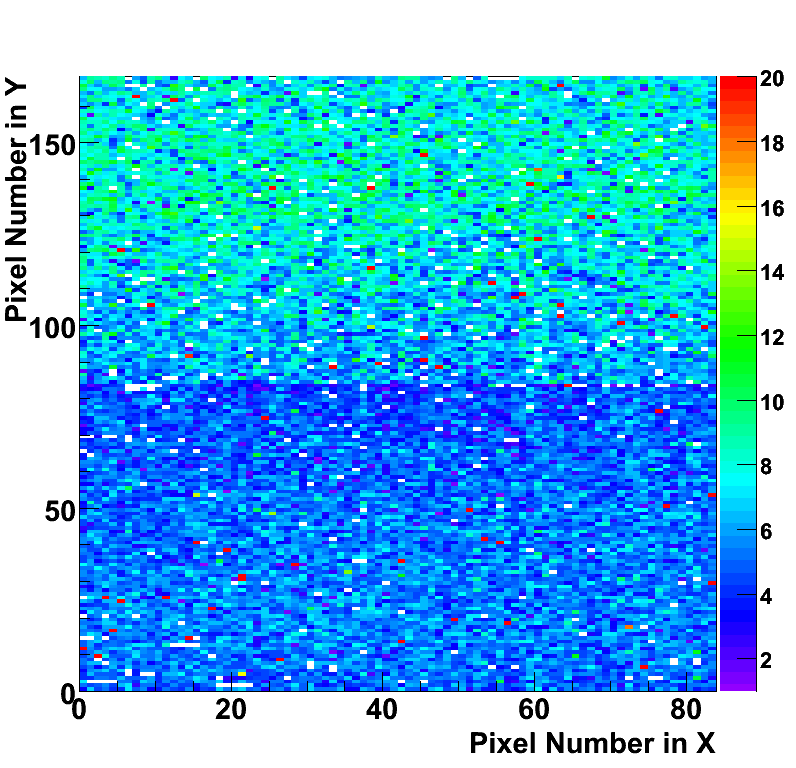}
\caption{\sl 
Two-dimensional map of  $^{55}$Fe peak position (left) and peak
width (right) for the two pre-shape quadrants of a typical sensor.
The lower half of each plot corresponds to quadrant~0
and the upper half to quadrant~1. The colour-coded scale
is in DAC threshold units (DTU) in both cases. Pixels for which no
peak was measured are shown as white. The faint vertical line in the 
centre of the left
plot corresponds to the pixels bordering the dead area between the two 
readout regions, which has no deep P-well implant.
}
\label{fig:pixel:ironmean}
\end{center}
\end{figure}

The variations in the position and width of the peak measured are shown
as a function of the position in the sensor in
figure~\ref{fig:pixel:ironmean}. The differences of the two quadrants
are clearly visible. However, within the quadrants,
similarly to figure~\ref{fig:pixel:twodpedestals},
there is no correlation between the gain or noise and the position in
the sensor.

For quadrant~0, the
average peak position relative to the pedestal was found to
be 117\,DTU. From section~\ref{sec:testp:calibration},
this corresponds to 34\% of the
1620\,$e^-$ of signal charge, i.e. $550\,e^-$,
so this calibrates 1\,DTU to be $4.7\,e^-$.
For quadrant~1, then 1\,DTU is equivalent to 3.3\,$e^-$.
This
can be used to interpret the noise measurements above.
The average noise values of 5\,DTU and 6\,DTU, for quadrants~0
and 1 respectively,
correspond to an equivalent in terms of
signal charge of $23\,e^-$ and $20\,e^-$.
Hence, the measured
bulk pixel noise is compatible with both the test pixel 
measurements and the predicted design value.

In terms of detection of MIP signals, then the typical signal
in the pixel hit by a charged particle is the plateau fraction
of around 30\% of the $1100\,e^-$ deposited, i.e. $350\,e^-$.
Hence, this corresponds to an expected typical signal/noise ratio 
for MIPs of around 15.

\subsection{Bulk laser signal response}
\label{sec:pixel:laser}
\begin{figure}[ht!]
\begin{center}
\includegraphics[width=6cm]{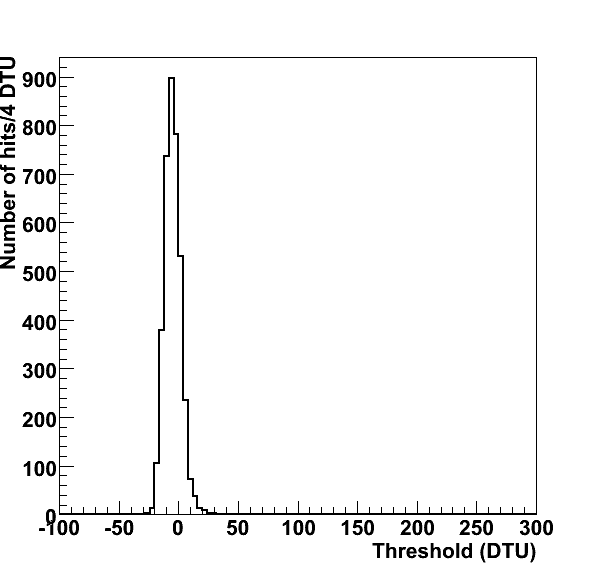}
\includegraphics[width=6cm]{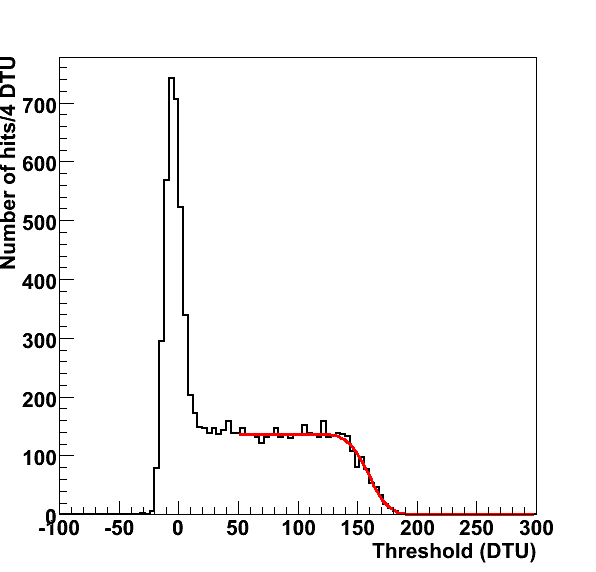}
\caption{\sl 
Typical threshold scan without (left) and with (right) the
laser powered on.
The right plot also shows the fit to the error function (red line), from
which the laser signal was determined to 
be 160\,DTU in this example.
}
\label{fig:pixel:laserscan}
\end{center}
\end{figure}

The sensor was
illuminated using a laser of wavelength 1064\,nm.
The laser enclosure contained a computer-controlled
moveable XY stage, on which the sensor PCB could be
mounted. This allowed the sensor to be moved in
two dimensions below the laser, allowing a large
number of measurements which are described in
this section.
For these measurements,
the signal was injected into the bulk
pixel area and the response measured using a threshold
scan. A typical response is shown in 
figure~\ref{fig:pixel:laserscan}, where the laser
signal is clearly seen. The actual size of the laser
signal was found by fitting the falling edge of the
response curve to an integrated Gaussian (an error function).
In terms of the integrated Gaussian parameters,
this gives a pseudo-analogue signal response estimate
from the mean and an estimate of the signal spread
from the width. The absolute intensity of the laser
was uncalibrated so the signal size for the
following measurements is arbitrary, although it is
similar to that expected for a MIP.

\begin{figure}[ht!]
\begin{center}
\includegraphics[width=6cm]{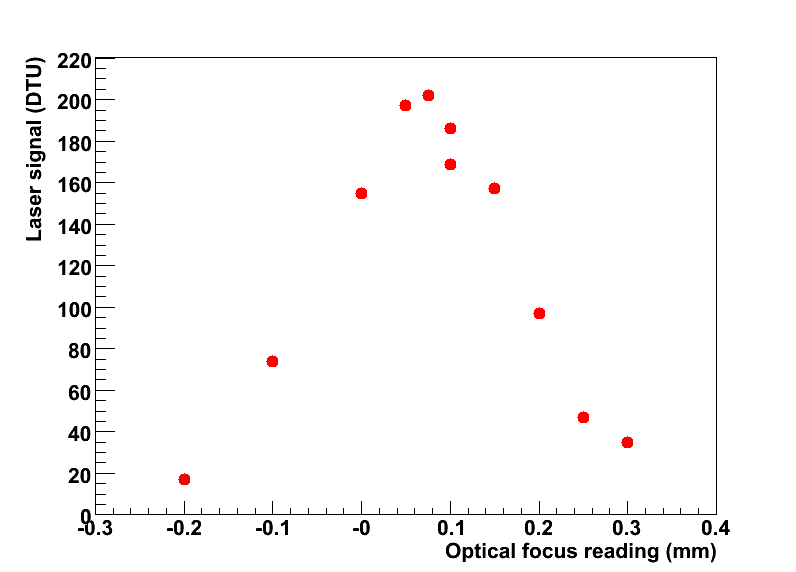}
\caption{\sl 
Laser signal size as a function of the focus depth
relative to the optical focus position. A positive
value means the focus position was moved into
the sensor.
}
\label{fig:pixel:focus}
\end{center}
\end{figure}

The laser must be focussed onto the epitaxial layer.
This was done by focussing the microscope optically on
the substrate surface and then recording the observed
laser signal size at various focus depths relative to
this. Figure~\ref{fig:pixel:focus} shows the results
of such a scan, where there is seen to be a strong 
dependence on the focus depth. The maximum signal was
found to be 60\,$\mu$m into the substrate relative
to the optical focus. In addition,
there is a range of only around $\pm 20\,\mu$m where the
signal size is effectively unchanged. This sets a
limit on the degree of tilt acceptable when performing
such laser measurements. A focus on pixels near one edge of the
sensor will only be within this tolerance for pixels near the
other edge if the sensor is flat to around 2\,mrad.
For this reason, 
the focus was checked when large movements
were made and, when possible, only small movements were
then performed around this point.

\begin{figure}[ht!]
\begin{center}
\includegraphics[width=6cm]{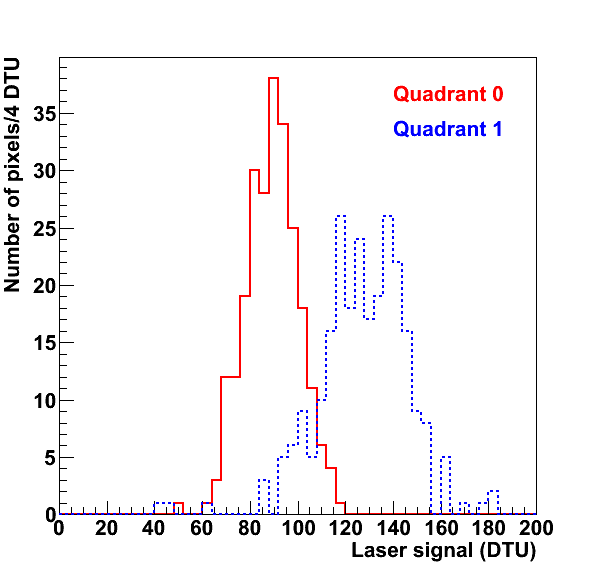}
\includegraphics[width=6cm]{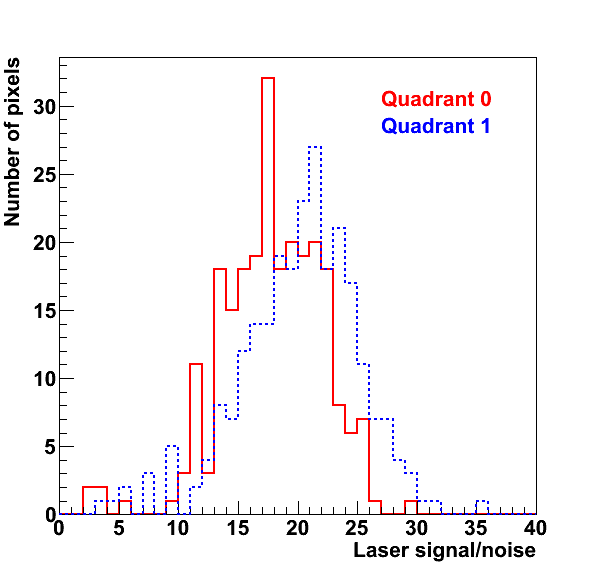}
\caption{\sl 
Laser signal (left) and signal/noise (right) for 
quadrants~0 (solid, red) and 1 (dashed, blue).
}
\label{fig:pixel:lasersignal}
\end{center}
\end{figure}

The laser system was used to determine the relative
gain of the pixels. By firing the laser into the centre
of each pixel in turn, then the signal size for each
was determined. Figure~\ref{fig:pixel:lasersignal}
shows the distribution of the observed laser signals and
signal/noise ratios for
a subset of pixels in the two pre-shape quadrants.
Comparison with figure~\ref{fig:pixel:irongain} shows
these results are consistent with those from the $^{55}$Fe
measurements. 
Again a clear difference in both signal size
(90\,DTU and 130\,DTU, respectively)
and signal/noise (18.4 and 20.6, respectively)
for quadrants~0 and 1 was observed, with quadrant~1 being
higher in both cases.
The signal size (i.e. gain) spread
for quadrants~0 and 1 was found to be 12\% and 14\% respectively,
which are slightly higher than, but consistent with, 
the $^{55}$Fe measurements.
Because of the better signal/noise performance,
only the quadrant~1 variant of the pre-shape pixels
was implemented in the next version of the sensor.

%

\begin{figure}[ht!]
\begin{center}
\includegraphics[width=7cm]{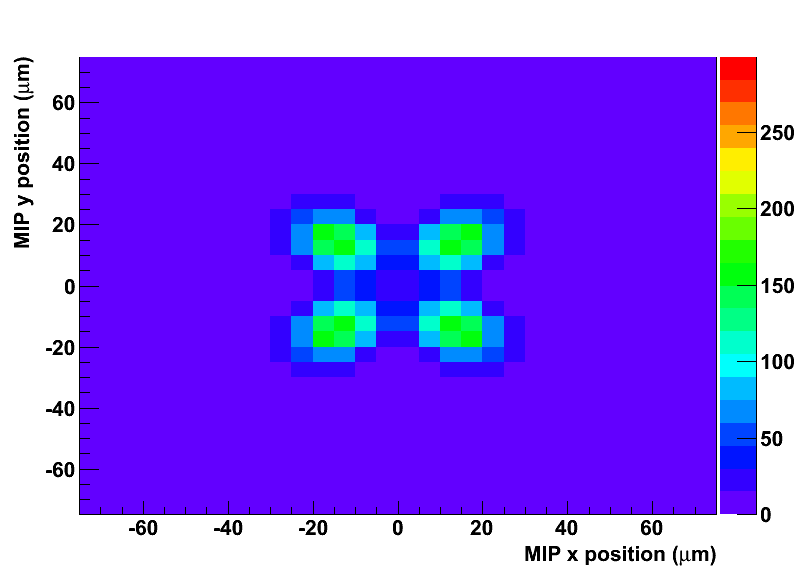}
\includegraphics[width=7cm]{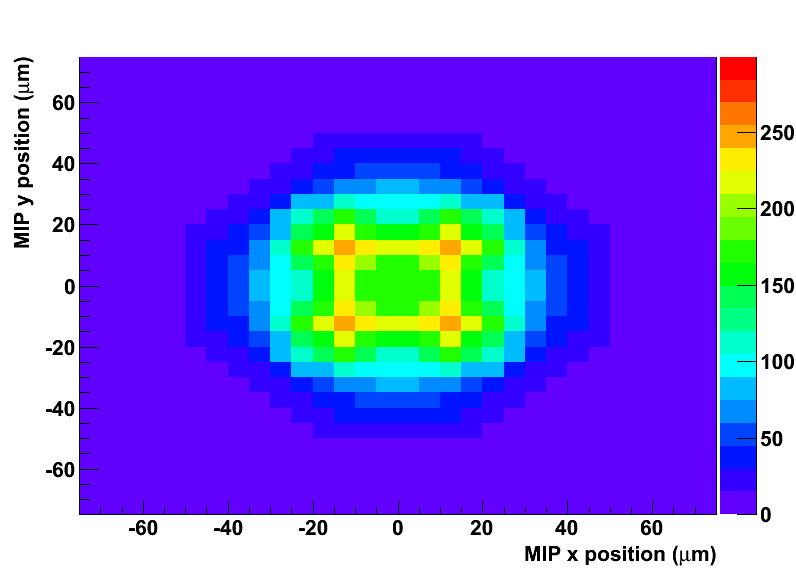}
\caption{\sl 
Charge seen as signal (in units of DTU) from laser
for no deep P-well (left)
and deep P-well (right), as a function of the laser impact position
relative to the centre of the pixel.
These data are from a sensor with a 12\,$\mu$m thick epitaxial layer.
}
\label{fig:pixel:bulklaser}
\end{center}
\end{figure}

The laser system was also used to measure the charge diffusion
so as to compare with the simulation described in
section~\ref{sec:simulation:sentaurus}. The laser was focussed to
give a illuminated spot size of less than 2\,$\mu$m and the
stage could be moved with a 1\,$\mu$m precision, so that 
the charge could be generated accurately to match the positions
of the 21 points shown in figure~\ref{fig:simulation:points}.
Because the laser was uncalibrated, the absolute fraction of
the laser signal observed could not be determined, 
so only the relative shape as a function of position is
relevant.
Figure~\ref{fig:pixel:bulklaser} shows the results of these
measurements using sensors with and without deep P-well
implants. Both sensors measured had a 12\,$\mu$m epitaxial
layer thickness.
Bearing in mind that the simulation (shown in
figure~\ref{fig:simulation:simcharge}) does not include the
effects of the laser spot size or electronics noise,
it is seen that the general trend of the data
is in quantitative agreement with the simulation
and that the deep P-well increases the signal size
significantly.
Overall, it is clear that
the sensor without deep P-well implants gives a
much smaller signal.

\section{Global sensor performance}
\label{sec:sensor}

Global aspects of the sensor performance were also studied and
are discussed below.

\subsection{Configuration load}
The per-pixel five-bit configuration data were loaded and
read back to check for bit errors. Each load transfered
141\,kbits into the sensor configuration registers. The
load and readback cycle was repeated 25,000 times and
the data written and readback were compared. No bits
were seen to have been corrupted in this process. This
sets an upper limit on the bit error rate of $1\times 10^{-9}$
at 90\% confidence level.

\subsection{Pickup between pixels}
\label{sec:sensor:pickup}

\begin{figure}[ht!]
\begin{center}
\includegraphics[width=6cm]{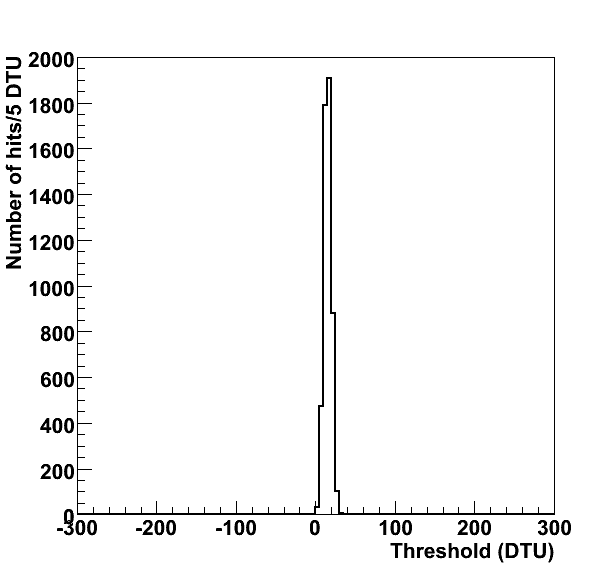}
\includegraphics[width=6cm]{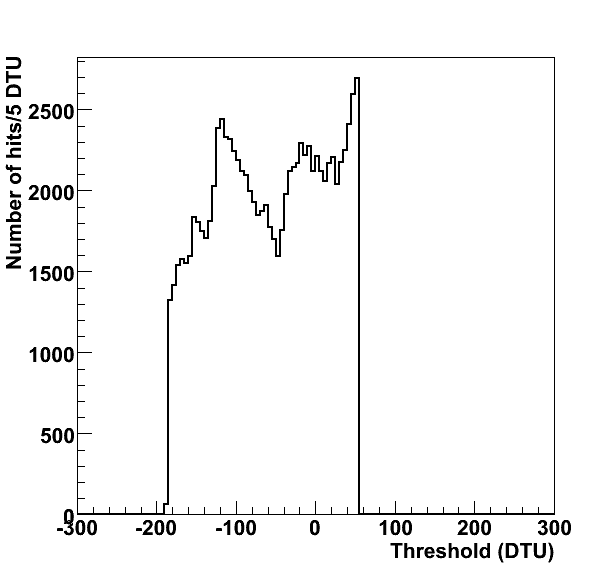}
\caption{\sl 
Effect of pickup on a typical pixel with only the 
pixel enabled (left) and with all other pixels enabled (right).
}
\label{fig:sensor:pickup}
\end{center}
\end{figure}

As mentioned in section~\ref{sec:pixel:peds}, pickup was
observed between pixels when large numbers fired at the
same time. For this reason,
all the results described in section~\ref{sec:pixel} were
done with only a small number of unmasked pixels.
Typical pickup behaviour is illustrated in
figure~\ref{fig:sensor:pickup} where the threshold
scan for a single pixel is shown with just that pixel
unmasked and also with all pixels unmasked.
It is clear the distribution when all pixels are enabled
is uncorrelated with the
pedestal of the actual pixel being studied;
it depends on the pedestals of the
other pixels firing. 
(The sensor was untrimmed in this case, and so the 
pedestals had a wide spread, as reflected in the width
of the distribution observed.)
The scan with pickup has a much
larger RMS and this has been used to characterise the
onset of pickup.

\begin{figure}[ht!]
\begin{center}
\includegraphics[width=7cm]{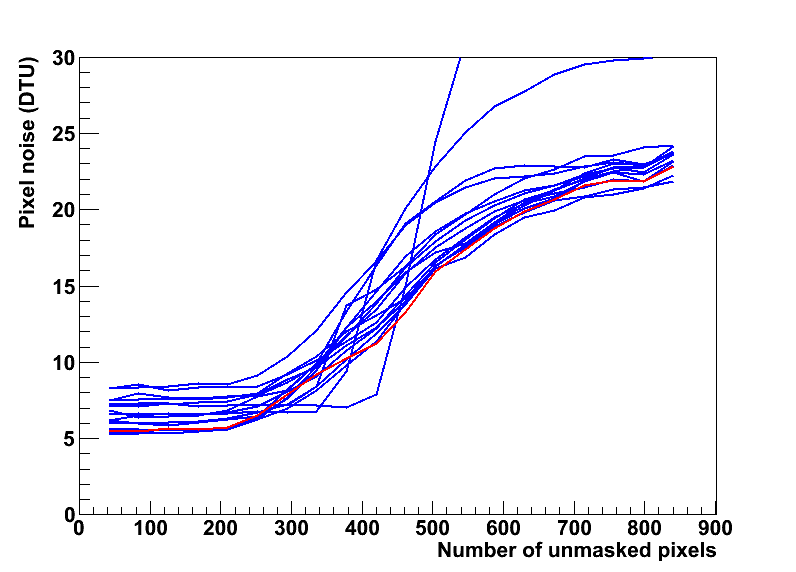}
\caption{\sl 
Onset of pickup, as measured by the RMS determined from a
threshold scan, as a function of the number of unmasked channels.
Each of the 16 lines corresponds to a single pixel, one
of which (in red) is on the opposite side of the sensor to the others.
It is seen that pickup starts on most pixels when around 300
pixels are unmasked and it is independent of the pixel position.
}
\label{fig:sensor:masking}
\end{center}
\end{figure}

Many studies to analyse this effect were performed. It was found
that the level of pickup depended on the number of pixels unmasked
but not their geometric positions relative to each other. One
example of these studies which shows this result clearly is shown
in figure~\ref{fig:sensor:masking}. In this study, a single pixel
on one edge of the sensor was unmasked. The RMS of this pixel
from a threshold scan was then determined as an increasing number
of pixels
were unmasked along the other edge of the sensor, i.e. at the
furthest point from the single pixel. It is seen that pickup on the
isolated pixel has the same behaviour as that of all the pixels
along the far edge. Hence, the pickup is seen to be unrelated to
position.

The pickup is not due to coupling directly between
monostables as this would not be affected by masking pixels.
It is thought that the pickup is caused by a power mesh
being shared between the pre-shape comparator and monostable
components. 
If a large number of pixels fire and
give hits in the monostables, the common power level
droops
and causes the comparators in other pixels to
fire. 
The power meshes to these two components have been
separated in the next version of the sensor.

It is important to note that this power coupling effect
does not affect normal operation of the sensor. 
The effect requires around 1\% of the pixels to
fire simultaneously. Hence,
it is only when doing sensor tests involving threshold
scans with low thresholds,
such as a pedestal determination, 
that large numbers of pixels fire simultaneously.
When operating
as a particle detector, then the threshold setting
is usually at least five times the noise value so
that the noise hit rate is much smaller than 1\%.
This means that only small numbers of pixels fire at
any one time and
hence the power droop is never large enough to cause spurious
hits.
In this mode, all pixels can be operated unmasked and
the sensor has high efficiency.

\subsection{Hit corruption}
The ``hit override'' mode described in
section~\ref{sec:design:storage} allows hit data to be
stored in the sensor memory without requiring any
pixels to be above threshold. In this mode, all seven
banks of pixels in each row of 42 pixels will store
data for every timestamp and hence the 19 data registers
will be filled within the first three timestamps.

This test mode results in predictable values for the
MUX address, timestamp and row address for every data word
stored. These can be compared with the values read out
to check for corruption. (Note, the six-bit hit data 
themselves are not predictable as noise can cause them
to be randomly set.) A low level of corruption, at
around 0.1\%, was seen for the MUX address and timestamp
bits, although the actual rate varied from sensor to sensor.
This was traced to a design fault; the signal to write to
the SRAM did not have a wide enough voltage range to guarantee
the value would be written correctly. The rate of corruption
was eliminated by applying an external voltage level of 2.7\,V
to increase the range and this was used for all the
results shown in this paper.
This design fault has been corrected in the next version
of the sensor.

\section{Conclusions}
\label{sec:conclusions}

We have produced a test CMOS sensor which demonstrates much
of the basic functionality
required for a binary ECAL. 
Test
results show one of the two pixel designs functions close to the expected
level.
Gain and noise values from both test and bulk pixels
are very similar to those predicted by
the design simulation. Also, the signal charge spread between
pixels is close to expectations from the pixel simulation.
As part of the sensor development, we worked with the
CMOS foundry to develop the INMAPS process, which enables
us to use a deep P-well implant to shield the signal
charge from the active PMOS circuit elements. Measurements
show this improves the signal collection by a factor of more
than three,
without affecting the gain or noise. This implant
makes the sensor viable.

This work will continue with a new sensor, TPAC~1.2,
which has been designed and fabricated.
The new sensor contains only the pre-shape quadrant 1
pixel design, which was shown to have the better signal-to-noise ratio,
modified to have the trim range extended from four to six bits.
It also implements several fixes, such as
the decoupling of the comparator and monostable power meshes
and correcting the SRAM memory write corruption.
Small adjustments have been made to the pixel layout to
improve the pixel uniformity for noise and gain. 
This sensor is the same
size as TPAC~1.0 and 1.1 and has the same number of pixels. It is 
I/O compatible and so can use the same sensor PCB
and DAQ system.
Results
from the new sensor are expected to be reported
in the near future.

\acknowledgments{This work was funded in part through a grant from the
Science and Technology Facilities Council (STFC), United
Kingdom.}


\begin{thebibliography}{99}

\bibitem{ref:introduction:ilc}
J.~Brau, Y.~Okada and N.~Walker (eds.),
\emph{International Linear Collider Reference Design Report}, (2007),
\href{http://www.linearcollider.org/rdr}
{\tt http://www.linearcollider.org/rdr}.
%
\bibitem{ref:introduction:calice}
CALICE Collaboration,
\href{https://twiki.cern.ch/twiki/bin/view/CALICE/WebHome}
{\tt https://twiki.cern.ch/twiki/bin/view/CALICE/WebHome}.
%
\bibitem{ref:introduction:prc}
C.~Adloff {\it et al.},
\emph{CALICE Report to the DESY Physics Research Committee},
(2011), arXiv:1105.0511.
%
\bibitem{ref:introduction:pfa}
J.-C.~Brient and H.~Videau, 
\emph{The calorimetry at the future e$^+$e$^-$ linear collider},
Proceedings of APS/DPF/DBP summer study on the future of particle physics, 
Snowmass, USA (2002), arXiv:hep-ex/0202004v1;
V.~L.~Morgunov,
\emph{Calorimetry Design with Energy-Flow Concept},
Proceedings of 10th International Conference on Calorimetry
in High Energy Physics (CALOR02), Pasadena, USA (2002);
M.~A.~Thomson, 
\emph{Particle Flow Calorimetry and the PandoraPFA Algorithm},
Nucl.\ Inst.\ Meth.\ {\bf A611} (2009) 25.
%
\bibitem{ref:introduction:decal}
J.~A.~Ballin {\it et al.}, 
\emph{A MAPS-based Digital Electromagnetic Calorimeter for the ILC},
Proceedings of 2007 International Linear Collider Workshop (LCWS07 and
ILC07), Hamburg, Germany (2007), arXiv:0709.1346.
%
\bibitem{ref:introduction:ichep}
P.~D.~Dauncey,
\emph{Performance of CMOS sensors for a digital electromagnetic calorimeter},
Proceedings of the 35th International Conference of High Energy Physics (ICHEP10),
Paris, France (2010),
\href{http://pos.sissa.it/archive/conferences/120/502/ICHEP 202010\_502.pdf}
{\tt http://pos.sissa.it/archive/conferences/120/502/\\
ICHEP\ 202010\_502.pdf}.
%
\bibitem{ref:design:history}
S.~Mendis, S.~E.~Kemeny and E.~R.~Fossum,
\emph{CMOS active pixel image sensor},
\emph{IEEE Trans. Electron Devices}, {\bf 41} (1994) 452;
E.~R.~Fossum, 
\emph{Active Pixel Sensors -- Are CCDs Dinosaurs?},
\emph{CCDs and Optical Sensors III, Proc. SPIE} {\bf 1900} (1993) 2.
%
\bibitem{ref:design:rt3}
R.~Turchetta {\it et al.},
\emph{A monolithic active pixel sensor for charged particle tracking and 
imaging using standard VLSI CMOS technology},
\emph{Nucl. Inst. Meth.} {\bf A458} (2001) 677;
G.~Deptuch {\it et al.}, 
\emph{Simulation and measurements of charge collection in monolithic 
active pixel sensors},
\emph{Nucl. Inst. Meth.} {\bf A465} (2001) 92.
%
\bibitem{ref:design:inmaps}
J.~A.~Ballin {\it et al.}, 
\emph{Monolithic Active Pixel Sensors (MAPS) in a quadruple well 
technology for nearly 100\% fill factor and full CMOS pixels},
\emph{Sensors} {\bf 8} (2008) 5336.
%
\bibitem{ref:design:eldo}
Mentor Graphics Eldo Integrated Circuit Simulation,
\href{http://www.mentor.com/products/ic_nanometer_design/analog-mixed-signal-verification/eldo/}
{\tt http://www.mentor.com/products/ic\_nanometer\_design/\\
analog-mixed-signal-verification/eldo/}.
%
\bibitem{ref:design:spectre}
Cadence Virtuoso Spectre Circuit Simulator,
\href{http://www.cadence.com/products/cic/spectre_circuit/pages/default.aspx}
{\tt http://www.cadence.com/products/cic/spectre\_circuit/pages/default.aspx}.
%
\bibitem{ref:design:sid}
SiD Detector Concept,
\href{http://SiliconDetector.org/}
{\tt http://SiliconDetector.org/}.
%
\bibitem{ref:design:ild}
ILD Detector Concept,
\href{http://www.ilcild.org/}
{\tt http://www.ilcild.org/}.
%
\bibitem{ref:simulation:sentaurus}
Sentaurus TCAD Process and Device Simulation Tools,
\href{http://www.synopsys.com/products/tcad/tcad.html}
{\tt http://www.synopsys.com/products/tcad/tcad.html}.
%
\bibitem{ref:pixel:noise}
A.~Papoulis and S.~U.~Pallai,
\emph{Probability, Random Variables and Stochastic Processes},
Section 11.4, 4$^{th}$ edition, McGraw Hill, 2002.

\end{thebibliography}
\end{document}